\documentclass[12pt]{iopart}
\usepackage{graphicx} 
\usepackage{psfig}
\begin{document}

\title[Cosmology with GRBs] {Gamma Ray Bursts as standard candles to constrain the
cosmological parameters}

\author{G. Ghirlanda, G. Ghisellini, C. Firmani}

\address{Osservatorio Astronomico Brera, Via Bianchi 46, I-23807 Merate (LC), IT}
\ead{giancarlo.ghirlanda@brera.inaf.it}

\begin{abstract}
Gamma Ray Bursts (GRBs) are among the most powerful sources in the
Universe: they emit up to $10^{54}$ erg in the hard X--ray band in few
tens of seconds.  The cosmological origin of GRBs has been confirmed
by several spectroscopic measurements of their redshifts, distributed
in the range $z\in(0.1,6.3)$. These two properties make GRBs very
appealing to investigate the far Universe.  Indeed, they can be used
to constrain the geometry of the present day universe and the nature
and evolution of Dark Energy by testing the cosmological models in a
redshift range hardly achievable by other cosmological probes.
Moreover, the use of GRBs as cosmological tools could unveil the
ionization history of the universe, the IGM properties and the
formation of massive stars in the early universe.  The energetics
implied by the observed fluences and redshifts span at least four
orders of magnitudes.  Therefore, at first sight, GRBs are all but
standard candles.  But there are correlations among some observed
quantities which allows us to know the total energy or the peak
luminosity emitted by a specific burst with a great accuracy.  Through
these correlations, GRBs becomes ``known" candles, and then they
become tools to constrain the cosmological parameters.  One of these
correlation is between the rest frame peak spectral energy $E_{\rm
peak}$ and the total energy emitted in $\gamma$--rays $E_{\gamma}$,
properly corrected for the collimation factor.  Another
correlation, discovered very recently,  relates the total GRB
luminosity $L_{\rm iso}$, its peak spectral energy $E_{\rm peak}$ and
a characteristic timescale $T_{0.45}$, related to the variability of
the prompt emission.  It is based only on prompt emission
properties, it is completely phenomenological, model independent and
assumption--free.  These correlations have been already used to find
constraints on $\Omega_{\rm M}$ and $\Omega_{\Lambda}$, which are
found to be consistent with the concordance model.   The present
limited sample of bursts and the lack of low redshift events,
necessary to calibrate the correlations used to standardize GRBs
energetics, makes the cosmological constraints obtained with GRBs
still large compared to those obtained with other cosmological probes
(e.g. SNIa or CMB).  However, the newly born field of GRB--cosmology
is very promising for the future.
\end{abstract}

\maketitle

\section{Introduction}

The present--day Universe and its past evolution are described in
terms of a series of free parameters which combined together make up
the cosmological model. One of the key objectives of observational
cosmology is to exploit the power of combining different observations
in order to test different cosmological models and find out the
``best'' one.  The standard model of cosmology can have up to about 20
parameters needed to describe the background space--time, the matter
content and the spectrum of the metric perturbations
(e.g. \cite{liddle,lahav} ). However, a ``minimal'' subset of 7
parameters can be used to construct a successful cosmological
model. In addition to the set of parameters describing the global
geometry and dynamics of the Universe (in terms of its curvature and
expansion rate - i.e.  $\Omega_{M}$, $\Omega_{b}$, $\Omega_{k}$ and
$h$), it became of greatest interest also the description of how the
structure were formed from the basic constituents (described in terms
of the perturbations amplitude $A$), the ionization history of the
universe since the era of decoupling ($\tau$) and the description of
how the galaxies trace the dark matter distribution on the largest
scales (described in terms of the bias parameter $b$).

The goal of observational cosmology is to use astronomical objects and
observations to derive the cosmological parameters (i.e. cosmological
test - e.g \cite{sanders,elizalde} ) once a cosmological model has
been defined by selecting a set of parameters (i.e. the model
selection problem - e.g. \cite{liddle,kunz}).

The Hubble constant $h=0.72\pm0.1$ has been recently measured by the
HST key project through the calibration of primary (Cepheid variables)
and secondary (SN type I, SN type II, Galaxies) distance indicators
within 600 Mpc. This measure has been confirmed by the CMB WMAP data
and by large scale structure measures
(\cite{spregel},\cite{spregel05}).

Supernovae Ia, through the luminosity distance test (see below), have
been used to constrain (combined with the evidence of a flat universe
from CMB data) $\Omega_{M}\simeq0.3$ and $\Omega_{\Lambda}\simeq0.7$
(e.g. \cite{knop,torny,riess,astier}).  One of the greatest
breakthrough of the cosmological use of SNIa is the evidence of the
recent re-acceleration of the universe (e.g.
\cite{riess98,perlmutter,riess00} see also \cite{filippenko} for a
review), interpreted as the effect of a still obscure form of energy
with negative pressure.

The CMB primary anisotropies bear the imprint of the physical
conditions at the epoch of matter--radiation decoupling and of several
effects such as the time-varying gravitational potential of structures
along its propagation path (the ISW effect), the gravitational lensing
and the scattering by the homogeneous ionized gas and by the
collapsed--ionized gas (the SZ effect). Among the greatest
breakthroughs of the CMB data analysis (\cite{spregel05}) it is worth
mentioning that (i) the universe is flat, at the 1\% level of
accuracy, (ii) $\tau\sim0.09\pm0.03$, (iii) the power spectrum of the
initial perturbations has the form of a scale invariant powerlaw, (iv)
a static dark energy model is preferred. In particular the combination
of the three year WMAP data with the Supernova Legacy data
(\cite{astier}) yields a significant constrain on the dark energy
equation of state, i.e. $w=-0.97\pm0.02$, favouring the $\Lambda CDM$
cosmological model.

The study of the power spectrum of large galaxy surveys allowed to
derive the baryon fraction $\Omega_{\rm b}/\Omega_{M}=0.185\pm0.046$
(\cite{cole}) or, independently from the CMB priors,
$\Omega_{M}=0.273\pm0.025$ (\cite{eisenstein}). Galaxy clustering
analysis has also been used to put constraints, within the CDM model,
on the neutrino mass (e.g. \cite{hu}).  On the other hand Galaxy
clusters, combined with constraints on the baryon density from
primordial nucleosynthesis, have been used to constrain $\Omega_{M}$
and $h$ (\cite{white}). Finally, strong (by clusters potentials) and
weak gravitational lensing have been used to measure the mass power
spectrum and derive constraints on $\sigma_{8}$ and $\Omega_{M}$
(e.g. \cite{refregier}).

\begin{figure}\begin{center}
\resizebox{13cm}{11cm}{\includegraphics{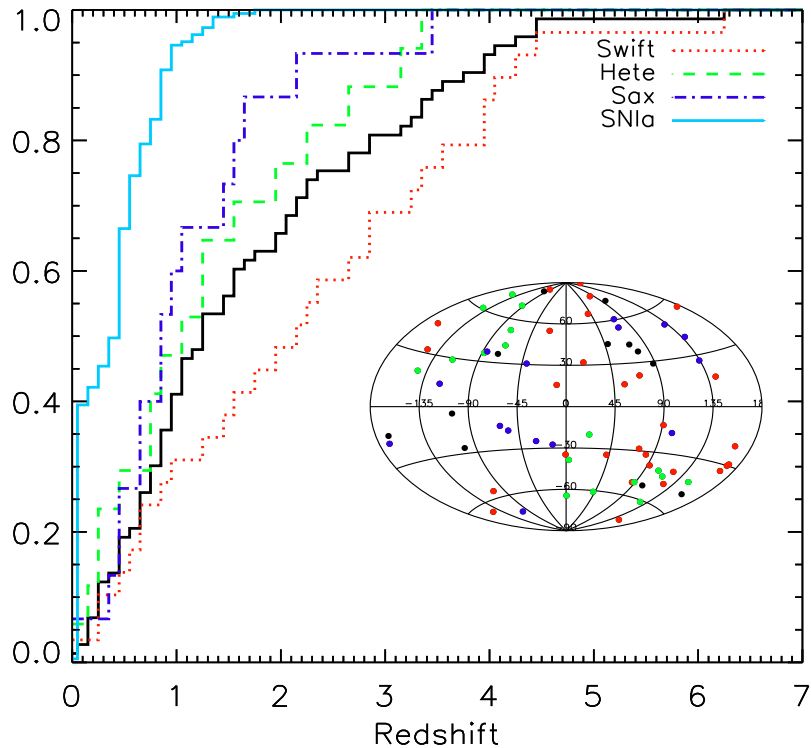}}
\caption{
Normalized redshift distribution of 79 long duration GRBs
updated to Jan 2006 (solid--black histogram). 
Also shown are the
distributions separated according to three different satellites that
detected most of these burst ($BeppoSAX$ [15 GRBs, dot-dashed blue
line], Hete-II [17 GRBs, dashed--green line] and Swift [29 GRBs,
dotted--red line]). The redshift distribution of the ``Gold'' SNIa
sample (156 objects) of \cite{riess} is also shown (solid--blue
line). The insert shows the sky distribution, in galactic
coordinates, of the present sample of GRBs with spectroscopically
measured redshifts (see http://www.mpe.mpg.de/~ jcg/grbgen.html for
reference).}
\label{z_dist}\end{center}
\end{figure}

Gamma Ray Bursts are very promising for cosmology for, at least, two
reasons: (i) they cover a very wide redshift range (see Fig.~\ref{z_dist},
solid black line) presently extending up to $z=6.29$ (\cite{kawai},
see \cite{web}) and (ii) GRBs are detected by space instruments in the
$\gamma$--ray band at energies $\ge$10 keV, i.e. their detection is
free from the typical limitations due to dust extinction in the
optical band.

The prospects of using GRBs as a new cosmological tool are exciting:
\begin{itemize}
\item 
similarly to QSO, GRBs might contribute to study the
distribution and the properties of matter observed in
absorption along the line of sight to these distant powerful sources
(e.g. \cite{fiore},\cite{lamb},\cite{lamb03},\cite{stratta}).
Moreover, given the high luminosity of the early afterglow and the
GRB transient nature, the absence of ``proximity effects''
(typical of QSO) makes GRBs powerful sources to study the metal
enrichment and ISM properties of their own hosts
(\cite{chen},\cite{fiore05}, \cite{berger} see also \cite{lazzati});
\item 
if GRBs correspond to the death of very massive stars, they
represent a unique probe of the initial mass function and of the star
formation of massive stars (\cite{bromm}) at very high redshifts;
\item  
GRBs can be potentially detected at any redshift: 
the spectroscopic study of GRBs will help to define the
properties of the IGM beyond the present QSO limits ($z\sim6$) and,
consequently to study the epoch(s) of re--ionization (e.g. \cite{inoue},
\cite{murakami})
\item 
the correlations between GRB spectral properties and collimation
corrected energetics (\cite{ghirlanda04},\cite{nava}), among prompt
observables (\cite{firmani06}) and prompt and afterglow observables
(\cite{liang}) have been shown to be powerful tools that
``standardize'' GRB energetics which can therefore be used to
constrain the universe dark matter and and energy content
($\Omega_{\rm M}$,$\Omega_{\Lambda}$) and the nature of dark energy
(\cite{ghirlanda04},\cite{firmani},\cite{xu},\cite{liang},\cite{lamb05},
\cite{ghirlanda06}).
\end{itemize}

GRBs are not {\it alternative} to SN Ia or other cosmological probes.
On the contrary, they are {\it complementary} to them, because
of their different redshift distribution and because any
evolutionary properties would likely to act differently on GRBs
than on SN Ia or other probes and because 
the joint use of different cosmological probes is the key to 
break the degeneracies in the found values of the cosmological parameters.


\subsection{The standard candle test}

The classical methods used to test the cosmological models are: (i)
the luminosity distance test (mainly applied to SN Ia), (ii) the
angular--size distance test (whose modern application consists in the
study of the CMB anisotropies) and (iii) the volume test based on
galaxy number count. Besides there are cosmological tests aimed at
testing structures' formation models such as the study of the
power spectrum of luminous matter or of the amplitude of the
present--day mass fluctuations.

For an object with known and fixed luminosity $L$, not evolving with
cosmic time, and measured flux $F$, we can define its luminosity
distance $D_{L}=(L/4\pi F)^{1/2}$ which is related to the radial
coordinate $r$ of the Friedman--Robertson--Walker metric by
$D_{L}=r/a(\tau)=r\,(1+z)$ (where $a(\tau)$ is the scale factor as a
function of the comoving time, $\tau=t\,H_{0}$).  Therefore, $D_{L}$
{\it depends} upon the expansion history and curvature of the universe
through the radial coordinate $r$. By measuring the flux of ``standard
candles'' as a function of redshift, $F(z)$, we can perform the
classical luminosity distance test by comparing $D_{L}$, obtained from
the flux measure, with $D_{L}(\bar{p})$ predicted by the cosmological
model, where $\bar{p}$ is a set of cosmological parameters
(e.g. $\Omega_{M}$, $\Omega_{\Lambda}$ and $h$).

This test has been widely used with SNIa (see \cite{perlmutter03} for
a recent review). The high peak luminosity (i.e. $\sim
10^{10}L_{\odot}$) of SNIa makes them detectable up $z>$1.  
However, SNIa, strictly--speaking,
are not standard candles: their absolute peak
magnitude varies by $\sim 0.5$ mag which corresponds to
a variation of 50\%--60\% in luminosity. In the early 1990s
(\cite{phillips},\cite{riess95},\cite{hamuy},\cite{goldhaber}) it was
discovered a tight correlation between the peak luminosity of SNIa and
the rate at which their luminosity decays in the post--peak phase. This
is known as the ``stretching--luminosity correlation'' and in its
simplest version\footnote{see also \cite{riess96} for and alternative
definition of the stretching--luminosity correlation based on
multi--color modeling of SNIa lightcurves.} is $M_{B}\simeq
0.8(\Delta m_{15} -1.1)-19.5$, where $M_{B}$ is the B--band absolute
magnitude and $\Delta m_{15}$ represents the decrease of the magnitude
from the time of the peak and 15 days later.
The application of this
correlation reduces the luminosity spread of SNIa to  
within
20\% and makes them usable as cosmological tools (\cite{hamuy}).

The situation with GRBs is remarkably similar. At first glance the
wide dispersion of their energetics prevented their use as
cosmological probes even when accounting for their ``jetted'' geometry
(Sec.~2). Several intrinsic correlations between temporal or spectral
properties and GRB isotropic energetics and luminosities (Sec.~3) have
been regarded as a possibility to make GRBs cosmological
tools. However, only by accounting for a third variable (which in the
standard model measures the GRB jet opening angle - Sec.~4), GRBs
became a new class of ``standard candles''  (or, better, ``known"
candles).  The constraints on the cosmological parameters obtained
by the luminosity distance test applied to GRBs are less severe than
what obtained with SNIa (Sec.~5). This is mainly due to the presently
still limited (i.e. only 19) number of GRBs with well determined
prompt and afterglow properties that can be used as standard candles (Sec.~9).

\section{GRB energetics and luminosities}

The  cumulative redshift distribution of long GRBs is compared in
Fig.~\ref{z_dist} with that of 156 SNIa (\cite{riess}).  We also
show the different redshift distributions obtained with the GRBs
detected by the three different satellites {\it Swift}, {\it Beppo}SAX
and {\it Hete--II}.  The better sensitivity and faster accurate
afterglow localization of {\it Swift} (\cite{gherels}), compared to
{\it Beppo}SAX and {\it Hete--II}, might account for the detection, on
average, of fainter X--ray and Optical afterglows and, therefore, for
systematically larger redshifts (\cite{berger05}).  It should also be
noted that due to its soft energy band (15--150 keV), {\it Swift}
might better detect and localize soft/dim long bursts. 
The insert of Fig.~\ref{z_dist} shows that the sky distribution of the
population of bursts with measured $z$ is consistent with that of long
GRBs detected by BATSE (see e.g. \cite{paciesas}).

\subsection{Isotropic energy and luminosity}

\begin{figure}\begin{center}
\resizebox{12cm}{10cm}{\includegraphics{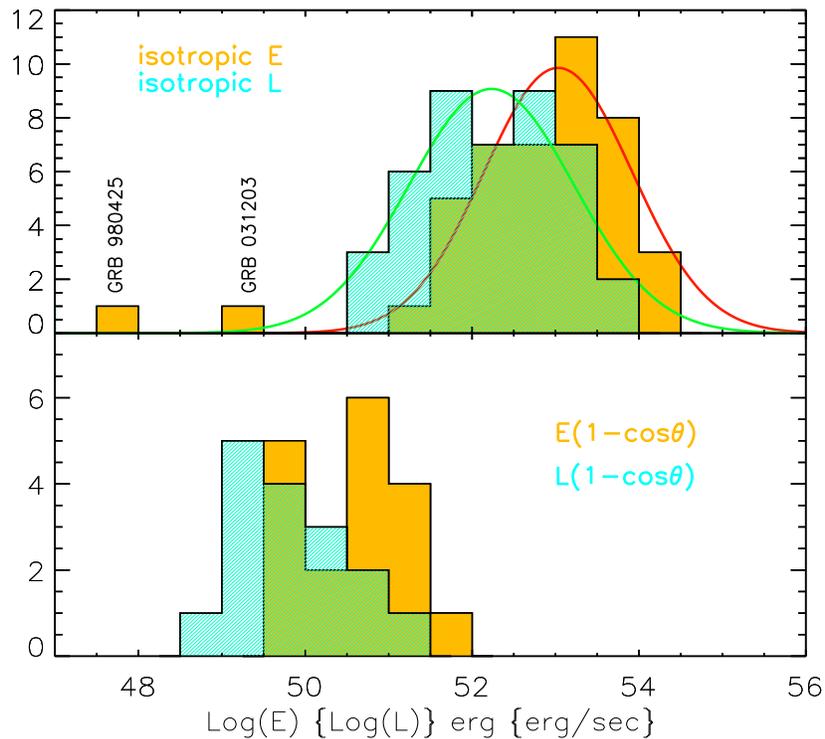}}
\caption[]{Upper panel: distributions of the isotropic
equivalent energy $E_{\rm iso}$ (orange--filled histogram) and of the
isotropic equivalent luminosity $L_{\rm iso}$ (blue-hatched
histogram). The sample comprises the 44 long GRBs with measured
redshifts and well determined spectral properties. The values of
$E_{\rm iso}$ are taken from \cite{amati06} while $L_{iso}$ is from
\cite{ghirlanda05} (updated with the most recent bursts). The solid
lines represent the Gaussian fit to these distributions.  The two
low--luminosity bursts (980425 and 031203) are also shown (see
\cite{ghirlanda04}, \cite{amati06}). Lower panel: distributions
of the collimation corrected energy (orange--filled histogram) and
luminosity (blue--hatched histogram). The samples comprises only the
19 GRBs with firm measurements of the jet break time from which the
jet opening angle could be computed (18 bursts from \cite{nava}
updated with GRB~051022, see \cite{ghirlanda06}).}
\label{edist}\end{center}
\end{figure}

The energy ($E$) and luminosity ($L$) of GRBs with measured redshifts
can be estimated through the burst observed fluence ($S$, i.e. the
flux integrated over the burst duration) and peak flux ($P$).  If GRBs
emit {\it isotropically}, the energy radiated during their prompt
phase is $E_{\rm iso}=4\pi D_{L}^2 S/(1+z)$ [where the term (1+z)
accounts for the cosmological time dilation effect] and the isotropic
luminosity is $L_{\rm iso}=4\pi D_{L}^2 P$.  Fig.~\ref{edist} (upper
panel) shows the distribution of $E_{\rm iso}$ (orange--filled
histogram) obtained with the most updated sample of 44 GRBs
(\cite{amati06}) with measured redshifts and measured spectral
properties (from which the bolometric fluence in the 1--10$^4$ keV
rest frame energy band can be computed). In the same plot we also show
the distribution of $L_{\rm iso}$ (blue--hatched histogram) which has
been obtained by updating the compilation of \cite{ghirlanda05} with
the most recent events.  Note that the  bolometric luminosity is
often computed by combining the peak flux (relative to the peak of the
GRB lightcurve) with the spectral data derived from the analysis
of the time integrated spectrum.  As discussed in \cite{ghirlanda05}
this is  strictly correct only if the GRB spectrum does not
evolve in time during the burst, contrary to what is typically
observed (\cite{preece}).

If the GRB energy and luminosity distributions of Fig.~\ref{edist}
(upper panel) are modeled with gaussian functions (solid--red line
and solid--green line, respectively), we find that the average
isotropic energy of GRBs is 
$\langle Log(E_{\rm iso}) \rangle=53.03\pm0.89$ 
while the average luminosity is 
$\langle Log(L_{\rm iso}) \rangle=52.23\pm0.10$ 
(1$\sigma$ uncertainty). 
Clearly, this large dispersion of $E_{\rm iso}$ and
$L_{\rm iso}$ prevents the application of the luminosity distance test
to GRBs.

%
\begin{figure}\begin{center}
\resizebox{12cm}{7cm}{\includegraphics{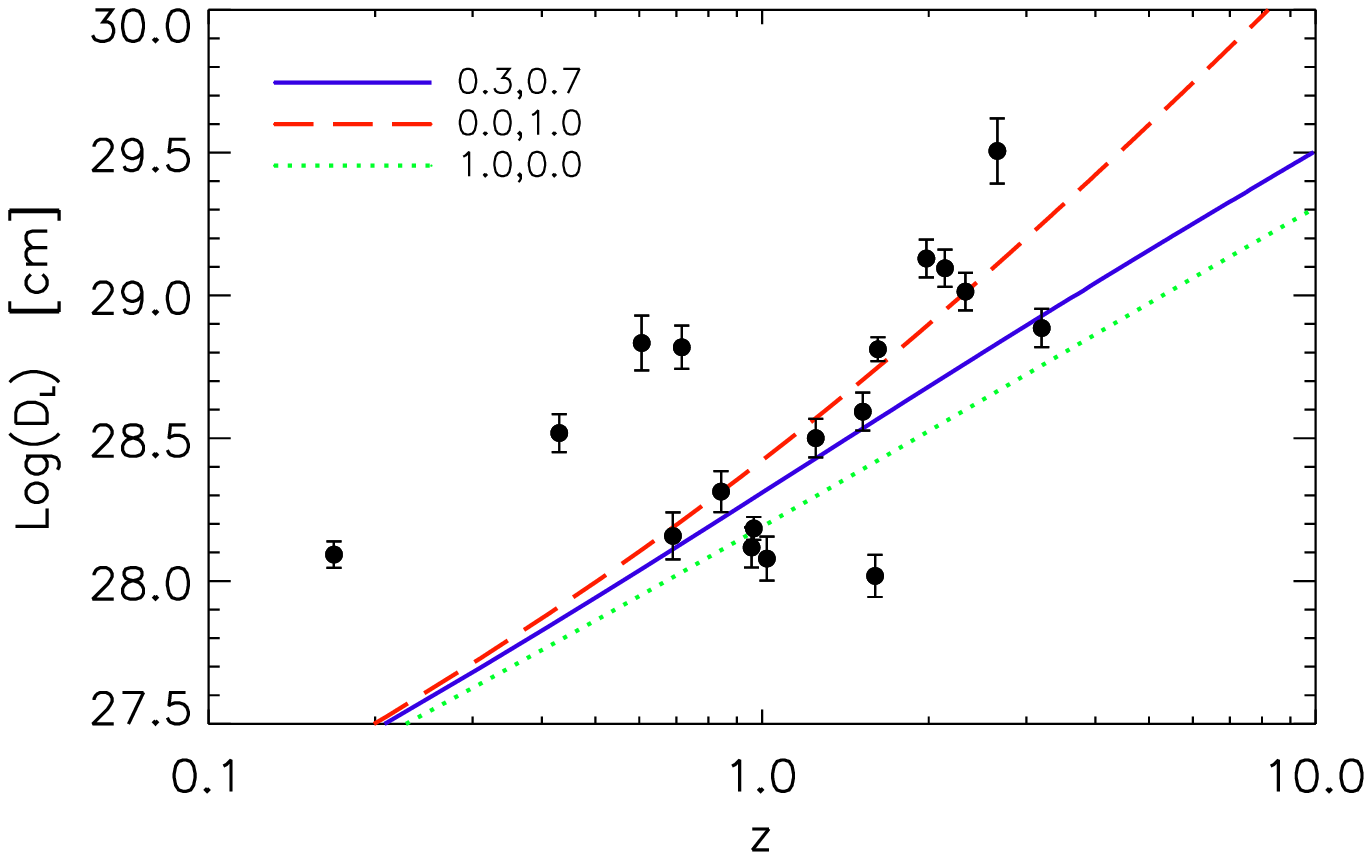}}
\caption{Hubble diagram for GRBs with measured redshifts and jet break
time assuming a unique value of the collimation corrected energy
$E_{\gamma}=8.8\times 10^{50}$ erg, i.e. the central value of its
distribution (blue--hatched histogram of Fig.\ref{hd}, lower
panel). The luminosity distance is obtained by solving the equation
$E_{\gamma}$=const=$E_{\rm iso}(1-\cos\theta)$.}
\label{hd}\end{center}
\end{figure}

\subsection{Collimation corrected energy and luminosity}

A considerable reduction of the dispersion of GRBs energetics has been
found by \cite{frail} (later confirmed by \cite{bloom}) when they are
corrected for the collimated geometry of these sources.

Theoretical considerations on the extreme GRB energetics under the
hypothesis of isotropic emission (\cite{waxman},\cite{fruchter}) led
to think that, similarly to other sources, also GRBs might be
characterized by a jet. In the standard scenario, the presence of
a jet (\cite{rohads}, \cite{sari}, \cite{dar} affects the afterglow
lightcurve which should present an achromatic break few days after the
burst (see Sec.~4 for more details).

Indeed, the observation of the afterglow lightcurves allows us to
estimate the jet opening angle $\theta_{\rm jet}$ from which the
collimation factor can be computed, i.e. $f=(1-\cos\theta_{\rm
jet})$. This geometric correction factor, applied to the isotropic
energies (\cite{frail}) led to a considerable reduction of the GRB
energetics and of their dispersion.  Fig.~\ref{edist} (lower panel)
shows the distributions of the collimation corrected energy
and peak luminosity ($E_{\gamma}=E_{\rm iso}\cdot f$ and $L_{\gamma}=L_{\rm
iso}\cdot f$, respectively) for the updated sample of bursts with
measured $\theta_{\rm jet}$. When compared with the parent
distributions of the isotropic quantities (upper panel of
Fig.~\ref{edist}) we note that, also accounting for the collimation
factor, both $E_{\gamma}$ and $L_{\gamma}$ are spread over $\sim$3
orders of magnitudes.

However, given the slightly lower dispersion of the collimation
corrected energies and luminosities (bottom panel of
Fig.~\ref{edist}) with respect to their isotropic equivalent (upper panel of Fig.~\ref{edist}), it is worth to verify if GRBs can
be effectively used as standard candles. The simplest test consists in
building the Hubble diagram (see also
\cite{bloom},\cite{ghirlanda04a}). By assuming the central value of
the collimation corrected energy, i.e. $E_{\gamma}=8.8\times 10^{50}$
erg, for all the GRBs with measured $z$ and $\theta_{\rm jet}$, it is
possible to derive their luminosity distance as if they were {\it
strictly speaking} standard candles, i.e. characterized by a unique
luminosity. By solving numerically the equation
$E_{\gamma}=8.8\times 10^{50}=E_{\rm iso}\cdot[1-\cos\theta_{\rm
jet}]$ (where also $\theta_{\rm jet}$ depends on $D_{\rm L}$ through
$E_{\rm iso}$, see Sec.~4) we derive the luminosity distance which is
plotted versus redshift in Fig.~\ref{hd}. Note that the scatter of the
data points is larger than the separation of different cosmological
models (lines in Fig.~\ref{hd}). This  simple tests shows that,
even if the collimation correction reduces the dispersion of GRB
energies, they cannot still be used as ``standard candles'' for the
luminosity distance test.

\section{The intrinsic correlations of GRBs}

For GRBs with known redshifts we can study their rest frame
properties. Although still based on a limited number (few tens)
of events, this analysis led to the discovery of several
correlations involving GRBs rest frame properties.  In the following
sections these correlations are summarized.

\subsection{The Lag--Luminosity correlation ($\tau$-$L_{\rm iso}$)}

The analysis of the light curves of GRBs observed by BATSE in 4 broad
 energy ranges (i.e.roughly 25-50, 50-100, 100-300 and $\ge$300 keV),
 led to the discovery of spectral lags: the emission in the  higher energy bands {\it precedes} that in the lower energy
 bands (\cite{norris}). The time lags typically range between
 0.01 and 0.5 sec (even few seconds lags have been observed -
 \cite{norris02}) and there is no evidence of any trend, within
 multipeaked GRBs, between the lags of the initial and the latest
 peaks (\cite{li}). It has been proposed that the lags are  a
 consequence of the spectral evolution (e.g.\cite{ford}), typically
 observed in GRBs (\cite{kocevsky}), and they have been interpreted as
 due to radiative cooling effects (\cite{bobbing}). Alternative
 interpretations invoke geometric (i.e. viewing angle) and
 hydrodynamic effects (\cite{ryde}, \cite{feng}) within the standard
 GRB model.

In particular, the analysis of the temporal properties of GRBs with
known redshifts revealed a tight correlation between their
spectral lags ($\tau$) and the luminosity ($L_{\rm iso}$)
(\cite{norris00}): more luminous events are also characterized
by shorter spectral lags as represented in Fig.\ref{lag_lum}
(left panel) where the original correlation $\tau\propto
L_{iso}^{-0.8}$, found with 6 GRBs, is reported (see also Tab.1).

Possible interpretations of the Lag-Luminosity correlation
include the effect of spectral softening of GRB spectra
(\cite{schaefer04}, \cite{daigne}) during the prompt due to
radiative cooling (\cite{crider}, \cite{liang96}) or a kinematic
origin due to the variation of the line--of--sight velocity in
different GRBs (\cite{salmonson}) or to the viewing angle of the jet
(\cite{ioka}).

Moreover, the $\tau-L_{iso}$ correlation has been used as
a pseudo-redshift indicator to estimate $z$ for a large population of
GRBs (\cite{band}) and also to study the GRB population properties
(i.e. jet opening angle, luminosity function and redshift
distribution) within a unifying picture (\cite{norris02}).

\subsection{The Variability--Luminosity correlation ($V$-$L_{\rm iso}$)}

The light curves of GRBs show several characteristic timescales. Since
their discovery, it was recognized that the $\gamma$--ray emission can
vary by several orders of magnitudes (i.e. from the peak to the
background level) on millisecond (or even lower (\cite{walker})
timescales (e.g \cite{lamb93}). Also the afterglow emission presents
some variability on timescales of a few seconds
(e.g.\cite{bersier},\cite{sato}). The analysis of large samples of
bursts also showed the existence of a correlation between the GRB
observer frame intensity and its variability (\cite{lamb93}). A short
timescale variability during the prompt emission of GRBs has been
considered as a strong argument against the external shock model,
favouring, instead, an internal shock origin for the burst high energy
emission (e.g. \cite{sari97}). In fact, the rapid variation of the
emission was interpreted as a signature of a discontinuous and rapidly
varying activity of the inner engine that drives the burst
(\cite{kobayashi}). This is hardly produced by a decelerating fireball
in the external medium. However, alternative scenarios propose
an external origin of the observed variability as due to the shock
formation by the interaction of the relativistically expanding
fireball and variable size ISM clouds (\cite{dermer}).

\begin{figure}[t]
\begin{minipage}[t]{8cm}
\resizebox{8cm}{7cm}{\includegraphics{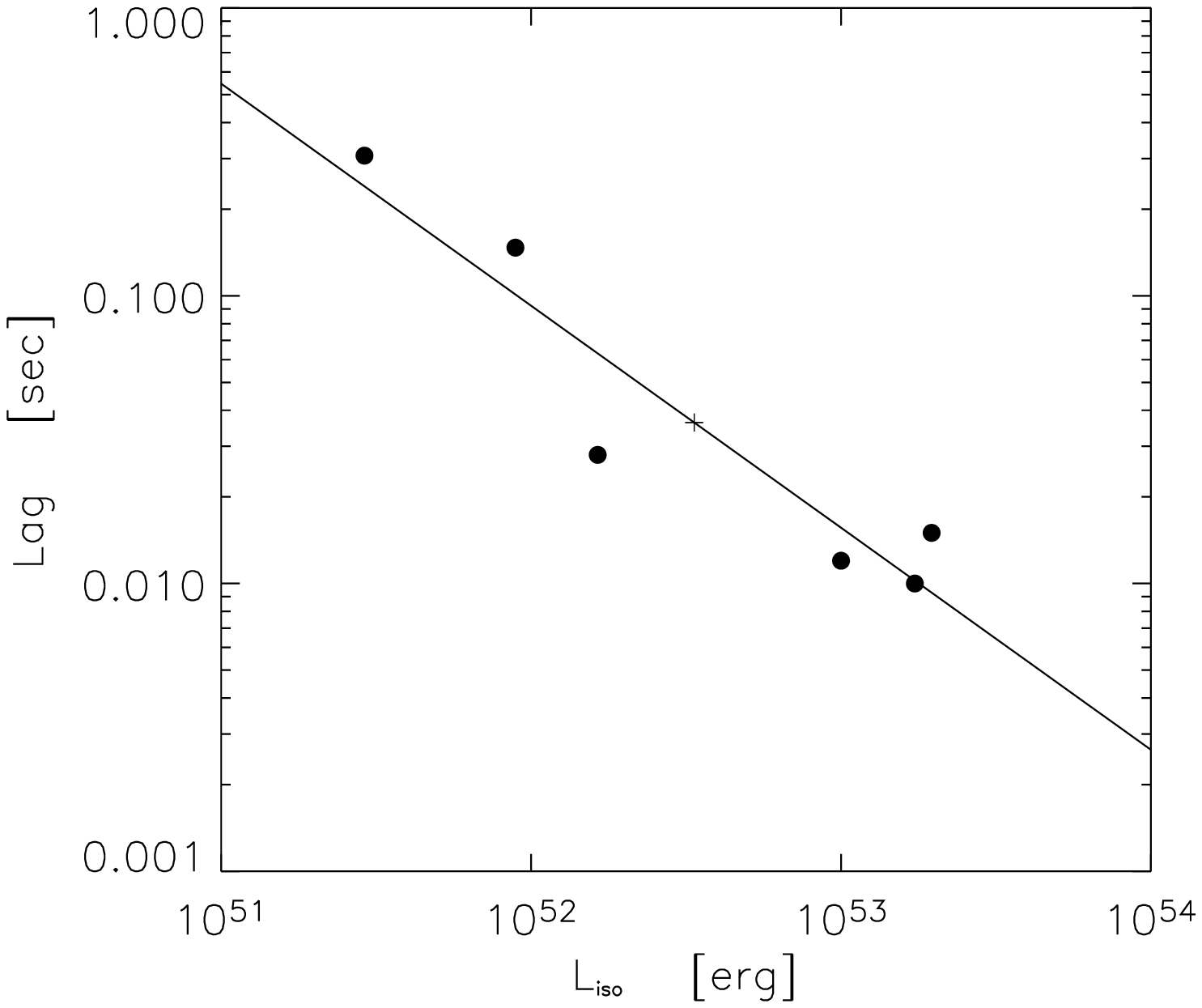}}
\end{minipage}\hfill
\begin{minipage}[t]{8cm}
\resizebox{8cm}{7cm}{\includegraphics{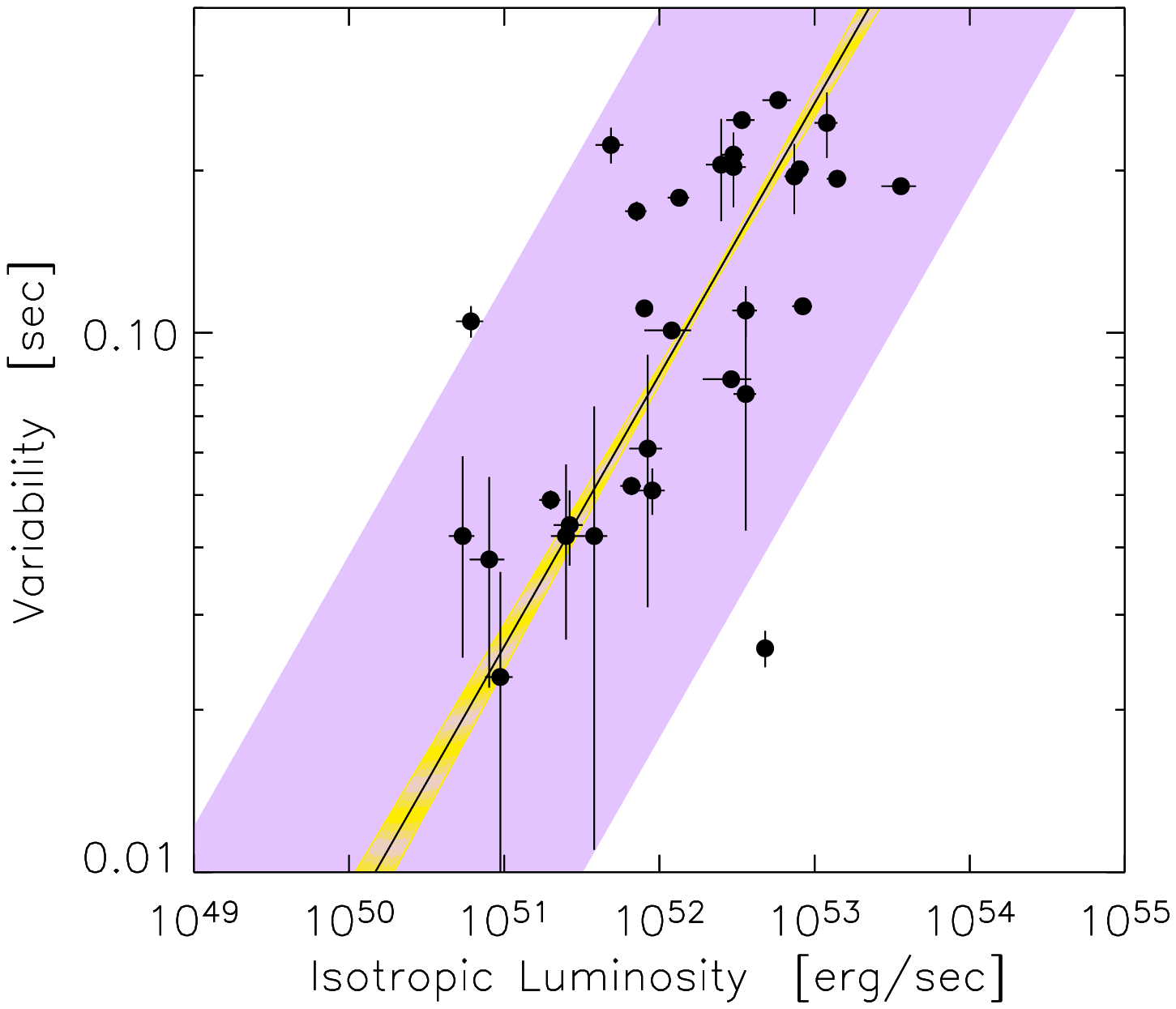}}
\end{minipage}
\caption[]{{\bf Left}: Lag--Luminosity correlation (data from
  \cite{norris}) for 6 GRBs with measured redshift. The best fit
  (solid line) is also shown. {\bf Right}: Variability--luminosity
  correlation for the sample of 31 GRB of \cite{guidorzi}. The
  violet--shaded region represents the 3$\sigma$ scatter of the data
  points around the best correlation (solid line).}
\label{lag_lum}
\end{figure}
\begin{figure}[htb]
\begin{minipage}[t]{8cm}
\resizebox{8cm}{7cm}{\includegraphics{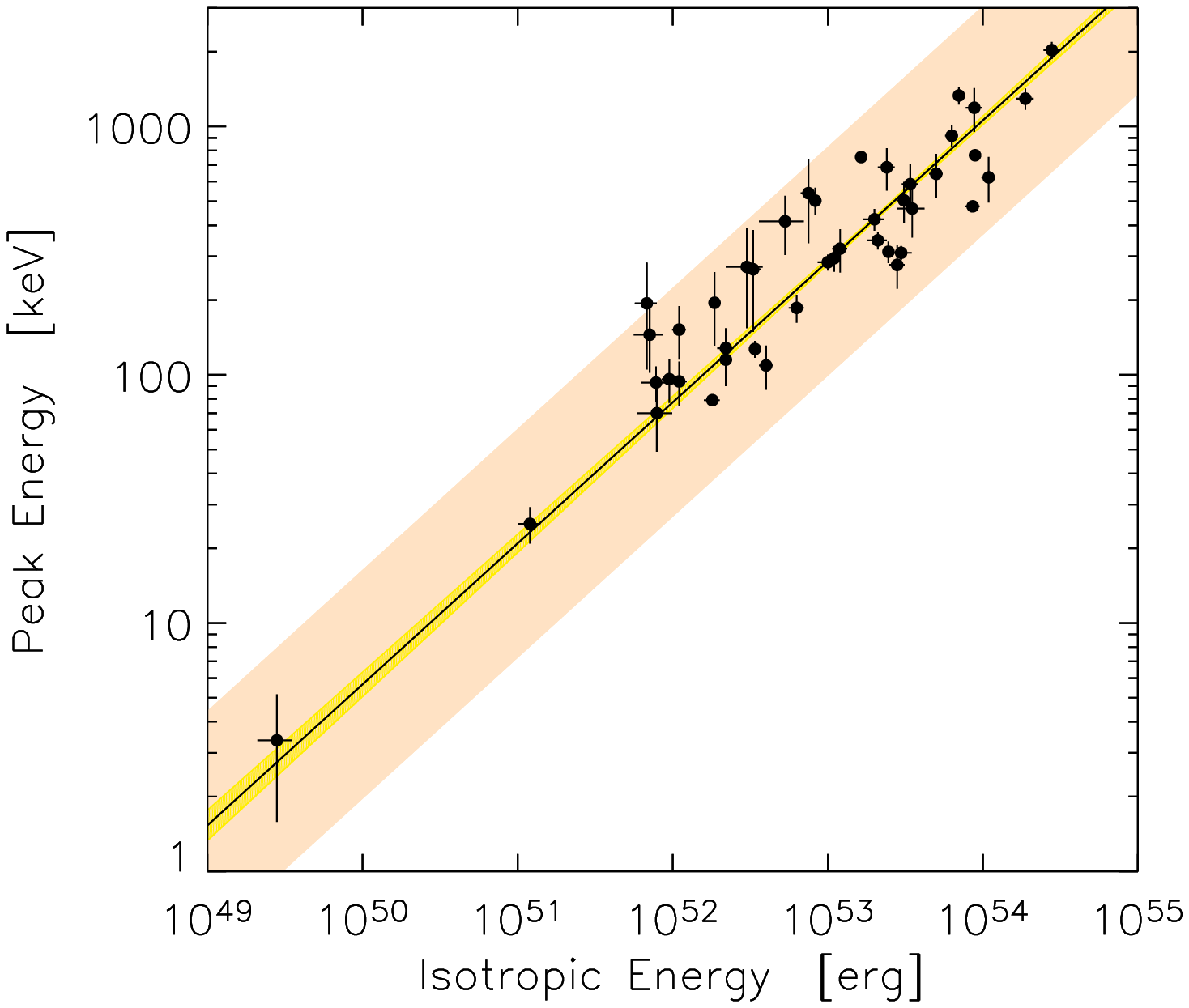}}
\end{minipage}\hfill
\begin{minipage}[t]{8cm}
\resizebox{8cm}{7cm}{\includegraphics{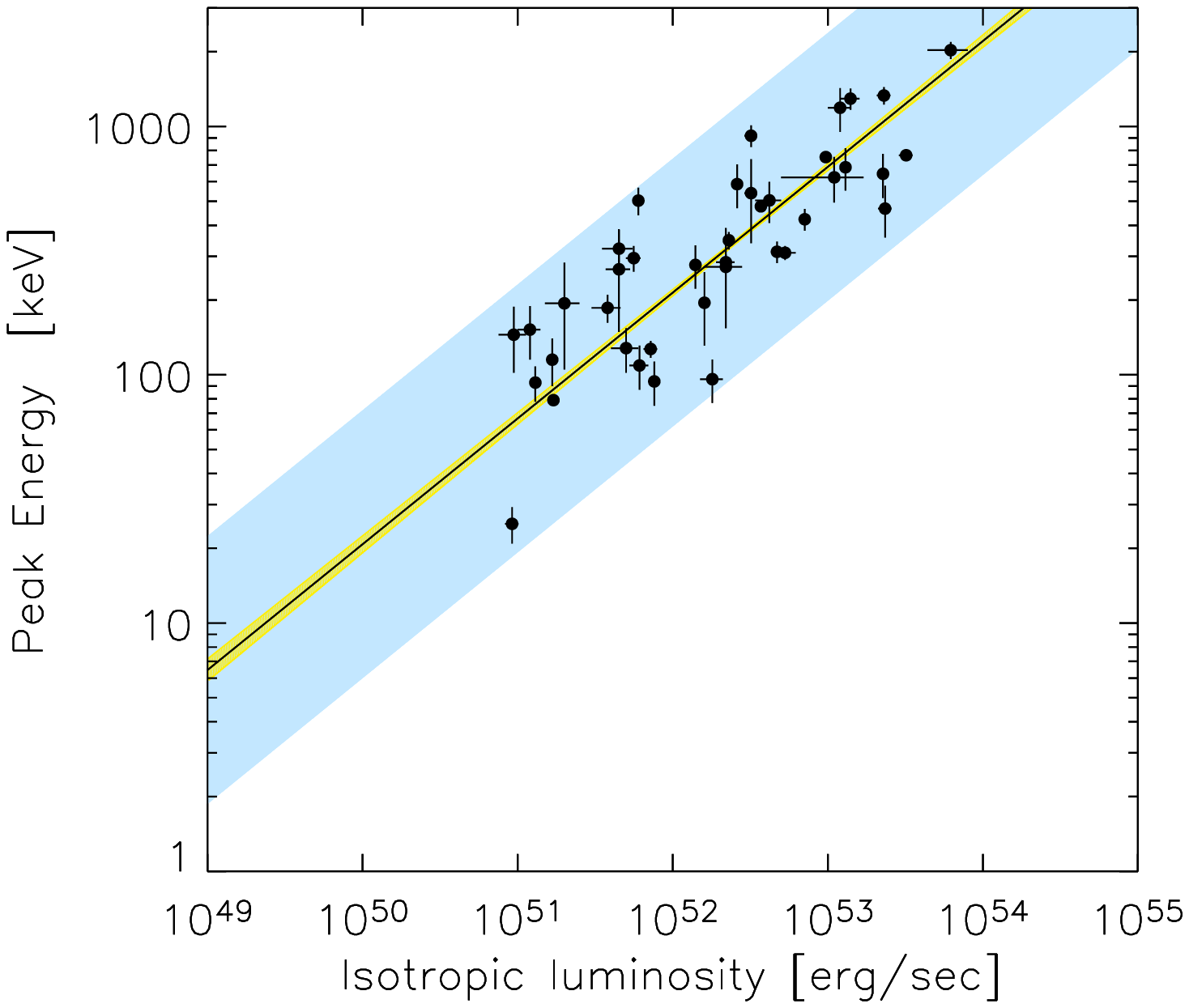}}
\end{minipage}
\caption[]{{\bf Left}: Peak energy--isotropic energy correlation. Data
are from \cite{amati06}.  The orange shaded region represents the
3$\sigma$ scatter of the data points around the correlation (solid
line). {\bf Right}: Peak energy--isotropic luminosity
correlation. Data (40 GRBs) are from \cite{ghirlanda05a} updated with
the most recent bursts.  The blue shaded region represents the
3$\sigma$ scatter of the data points around the correlation (solid
line).}
\label{edist1}
\end{figure}

\begin{table}\begin{center}
\caption{\label{tabella}GRB intrinsic correlations involving isotropic
quantities. N: number of GRBs used to find the correlation. The rank
correlation coefficient $r_s$ and its chance probability $P$ are
given. The best fit parameters of the powerlaw model (normalization
$q$ and slope $\alpha$), with their 1$\sigma$ uncertainty, are also
reported together with the $\chi^{2}$(dof).}  {\scriptsize
\begin{tabular}{@{}lllllll}
\br Isotropic Correlations & N & $r_{s}$ & $P$ & $q$ & $\alpha$ &
$\chi^{2}$(dof) \\ \hline Lag-Luminosity & 6 & -0.82 & 0.06 &
39.09$\pm$7.2 & -0.77$\pm$0.13 & ...  \\ $Log(\tau_{\rm lag})= q +
\alpha Log(L_{\rm iso})$ & & & & & & \\ Variability-luminosity& 31 &
0.61 & 8$\times10^{-4}$ & -27.3$\pm$1.3 & 0.5$\pm$0.02 & 1252(29) \\
$Log(V)= q + \alpha Log(L_{\rm iso})$ & & & & & & \\ Peak
energy-Isotropic energy& 43 & 0.89 & $10^{-15}$ & -27.64$\pm$0.65 &
0.56$\pm$0.01 & 493(41) \\ $Log(E_{\rm peak})= q + \alpha Log(E_{\rm
iso})$ & & & & & & \\ Peak energy-Isotropic luminosity& 40 & 0.86 &
$10^{-11}$ & -27.02$\pm$0.56 & 0.50$\pm$0.01 & 454(38) \\ $Log(E_{\rm
peak})= q + \alpha Log(L_{\rm iso})$ & & & & & & \\ \br
\end{tabular}}\end{center}
\end{table}

Fenimore \& Ramirez--Ruiz (\cite{fenimore}) and Reichart et
al. (\cite{reichart}) found a correlation between GRB luminosities
($L_{\rm iso}$) and their variability ($V$): more luminous bursts have
a more variable light curve. The $V-L_{\rm iso}$ correlation has been
recently updated (\cite{guidorzi}) with a sample of 31 GRBs with
measured redshifts. This correlation has also been tested
(\cite{guidorzi05}) with a large sample of 551 GRBs with only a
pseudo-redshift estimate (from the lag--luminosity correlation -
\cite{band}). An even tighter correlation (i.e. with a reduction of a
factor 3 of its scatter) has been derived (\cite{xin}) by slightly
modifying the definition of the variability first proposed by
\cite{reichart}. In Fig.\ref{lag_lum} we show the data points (from
\cite{guidorzi}) which show the existence of a statistically
significant correlation (with rank correlation coefficient
$r_{s}=-0.6$ and chance correlation probability
$P=8\times10^{-4}$). We fitted this correlation by accounting for the
errors on both coordinates and the best fit coefficients are reported
in Tab.1. The scatter of the data points around this correlation
(computed perpendicular to the correlation itself) can be modeled with
a gaussian with $\sigma=0.24$. As also shown in Fig.\ref{lag_lum}
(right panel) such a large scatter is responsible for a statistically
poor fit, as shown by the resulting $\chi^2$=1252/29. However, as
discussed in \cite{reichart05} when the sample variance (see also
\cite{reichart}, \cite{dagostini} ) is taken into account
($\sigma_{logV} = 0.20\pm0.04$) the correlation is $L\propto
V^{3.4(+0.9,-0.6)}$ with a reduced $\chi^{2}_{r}\sim 1$.

\subsection{The spectral peak energy--isotropic energy correlation ($E_{\rm peak}$-$E_{\rm iso}$)}

It has been shown (\cite{lloyd}) that GRBs with highly variable light
curves have larger $\nu F_{\nu}$ spectral peak energies. The existence
of a $V$-$E_{\rm peak}$ correlation and of the variability--luminosity
correlation ($V$-$L_{\rm iso}$) implies that the rest frame GRB peak
energy $E_{\rm peak}$ is correlated with the intrinsic luminosity of
the burst.  Amati et al. (\cite{amati}) analyzed the spectra of 12
$BeppoSAX$ GRBs with spectroscopically measured redshifts and found
that the isotropic--equivalent energy $E_{iso}$, emitted during the
prompt phase, is correlated with the rest--frame peak energy of the
$\gamma$--ray spectrum $E_{\rm peak}=E_{\rm peak}^{\rm
obs}(1+z)$. Such a correlation was later confirmed with GRBs detected
by BATSE, {\it Hete-II} and the {\it IPN} satellites and extended with
X-ray Flashes (XRF) towards the low end of the $E_{\rm peak}$
distribution
(\cite{lamb04},\cite{ghirlanda04},\cite{sakamoto}). Recently the
$E_{\rm peak}$-$E_{\rm iso}$ correlation has been updated with a
sample of 43 GRBs (comprising also 2 XRF) with firm estimates of $z$
and of the spectral properties (\cite{amati06}). In Fig.~\ref{edist1}
we show the $E_{\rm peak}$--$E_{\rm iso}$ correlation with the most
updated sample of GRBs reported in \cite{amati06}. The best fit
correlation parameters are reported in Tab.1. The scatter of the data
points around the $E_{\rm peak}$-$E_{\rm iso}$ correlation can be
modeled with a Gaussian of $\sigma\sim 0.15$.

The theoretical interpretations, proposed so far, of the $E_{\rm
peak}$--$E_{\rm iso}$ correlation ascribe it to geometrical effects
due to the jet viewing angle with respect to a ring--shaped emission
region (\cite{eichler},\cite{levinson}) or with respect to a multiple
sub-jet model structure which also accounts for the extension of the
above correlation to the X-ray Rich (XRR) and XRF classes
(\cite{yamazaki},\cite{toma}). An alternative explanation of the
$E_{\rm peak}$--$E_{\rm iso}$ correlation is related to the
dissipative mechanism responsible for the prompt emission
(\cite{rees}): if the peak spectral energy is interpreted as the
fireball photospheric thermal emission comptonized by a dissipation
mechanism (e.g. magnetic reconnection or internal shock) taking place
below the transparency radius, the observed correlation can be
reproduced.

\subsection{The peak spectral energy--isotropic luminosity correlation ($E_{\rm peak}$-$L_{\rm iso}$)}

A correlation between $E_{\rm peak}$ and the isotropic luminosity
$L_{\rm iso}$ (i.e. $E_{\rm peak}\propto L_{\rm iso}^{0.5}$) has been
discovered (\cite{yonetoku}) with a sample of 16 GRBs. A larger sample
of 25 GRBs confirmed this correlation (\cite{ghirlanda05}), although
with a slightly larger scatter of the data points. We have
further updated this sample to 40 GRBs with known redshifts and
Fig.~\ref{edist1} (right panel) shows the $E_{\rm
peak}$--$L_{\rm iso}$ correlation found with this sample. In
Tab.1 are reported the fit results of the $E_{\rm peak}-L_{\rm iso}$
correlation. The scatter of the data points around this correlation
(see Fig.\ref{scat}) can be modeled with a Gaussian of $\sigma\simeq
0.2$. Also in this case, as discussed for the $E_{\rm peak}-E_{\rm
iso}$ (\cite{amati06}) the resulting reduced $\chi^{2}$ is quite large
unless accounting for a sample variance of the order of $\sim 0.13$.

As discussed in \cite{ghirlanda05b}, the luminosity $L_{\rm iso}$ is
defined by combining the time--integrated spectrum of the burst with
its peak flux (also $E_{\rm peak}$ is derived using the time
integrated spectrum). This assumes that the time--integrated spectrum
is also representative of the peak spectrum. However, it has been
shown that GRBs are characterized by a considerable spectral evolution
(e.g. \cite{ford}). If the peak luminosity is derived only by
considering the spectrum integrated over a small time interval ($\sim
1$ sec) centered around the peak of the burst light curve , we find a
larger dispersion of the $E_{\rm peak}$-$E_{\rm iso}$
correlation  (see
\cite{ghirlanda05b}). This suggests that, in general, the time
averaged quantities (i.e. the peak energy of the time integrated
spectrum and the ``peak--averaged'' luminosity) are better correlated
than the ``time--resolved'' quantities.

Schaefer (\cite{schaefer03}) combined the Lag--Luminosity and the
Variability--Luminosity relations of 9 GRBs to build their Hubble
diagram. He showed that GRBs might be powerful cosmological tools.
However, some notes of caution should be addressed when using the
correlations presented in Tab.1 for cosmographic purposes.  In fact,
one fundamental condition is that there exist at least one cosmology
in which these correlations give a good fit, i.e. a reduced
$\chi^2_{red}\simeq 1$ and, unless accounting for the sample variance,
this is not the case for the correlations presented so far.

\subsection{The isotropic luminosity--peak energy--high signal timescale correlation ($L_{\rm iso}$--$E_{\rm peak}$--$T_{0.45}$) }

\begin{figure}[t]
\begin{center}
\resizebox{13cm}{13cm}{\includegraphics{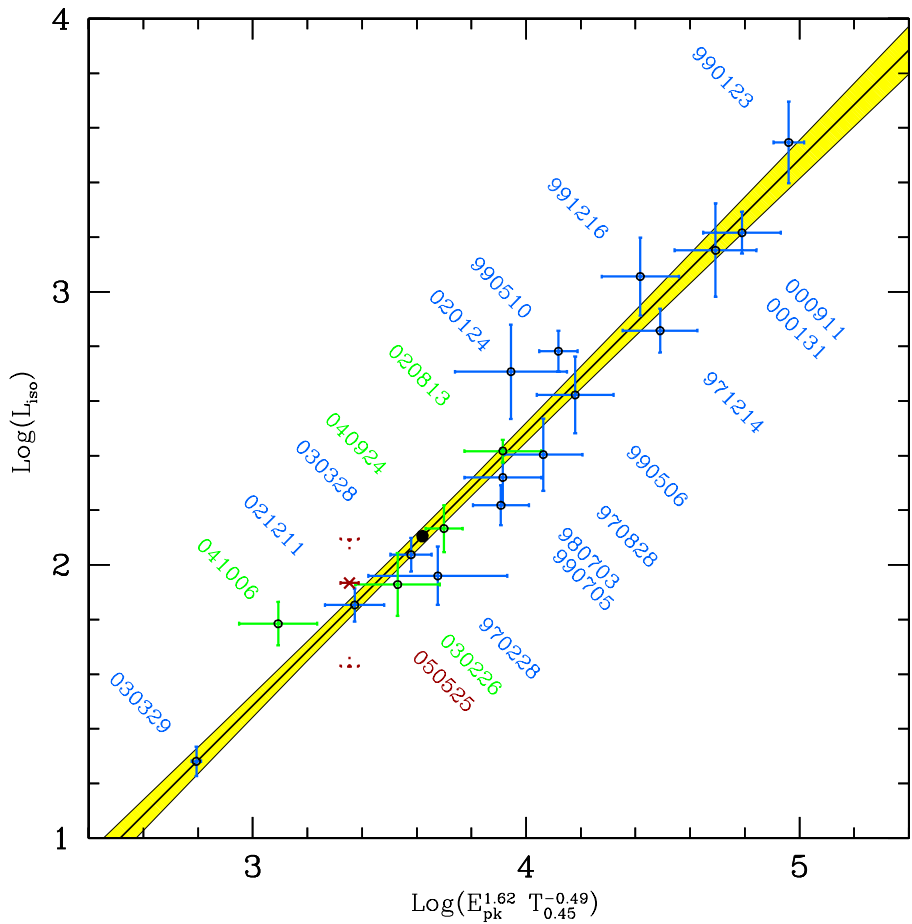}}
\caption{Multi variable linear regression analysis of the $L_{\rm
iso}$-$E_{\rm peak}$-$T_{0.45}$ correlation based on 19 GRBs. The open
circles represent the 19 bursts used for the fit (solid line). The
solid filled region represents the 1$\sigma$ uncertainty on the best
fit (see Eq.~1). The fit is performed in the centroid defined by the
data points (solid circle) where the errors on the best fit parameters
are uncorrelated. GRB~050525 was not used for the multi--variable
regression analysis due to its uncertain luminosity.}\label{firma}
\end{center}
\end{figure}

A new correlation recently discovered by Firmani et
al. (\cite{firmani06}) relates three observables of the GRB prompt
emission. These are the isotropic luminosity $L_{\rm iso}$, the rest
frame peak energy $E_{\rm peak}$ and the rest frame ``high signal''
timescale $T_{0.45}$. The latter is a parameter which has been
previously used to characterize the GRB variability
(e.g. \cite{reichart}) and represents the time spanned the brightest
45\% of the total counts above the background. Through the analysis of
19 GRBs, for which $L_{\rm iso}$ $E_{\rm peak}$ and $T_{0.45}$ could
be derived, \cite{firmani06} found that $L_{\rm iso}\propto E_{\rm
peak}^{1.62}\cdot T_{0.45}^{-0.49}$ with a very small scatter. This
correlation is presented in Fig.\ref{firma}.

The $L_{\rm iso}-E_{\rm peak}-T_{0.45}$ correlation is based on
prompt emission properties only and it has some interesting
consequences: (i) it represents a new powerful (redshift) indicator
for GRBs without measured redshifts, which could be computed only from
the prompt emission data (spectrum and light curve); (ii) it
represents a new powerful cosmological tool (\cite{firmani06a} - see
Sec.~6.2) which is model independent (differently from the
$E_{\gamma}-E_{\rm peak}$ correlation (see Sec.~5) which relies
on the standard GRB jet model); (iii) it is ``Lorentz invariant'' for
normal fireballs, i.e. when the jet opening angle is $\theta_{\rm
jet}>1/\Gamma$. In this case, in fact, the luminosity scales as
$\delta^{-2}$ when it is transformed from the rest to the comoving
frame and the doppler factor cancels out with that of the peak energy
and of $T_{0.45}$ - see \cite{firmani06}.

\section{The third observable: the jet break time ($t_{\rm break}$)}

\begin{figure}[t]
\begin{center}
\resizebox{12cm}{12cm}{\includegraphics{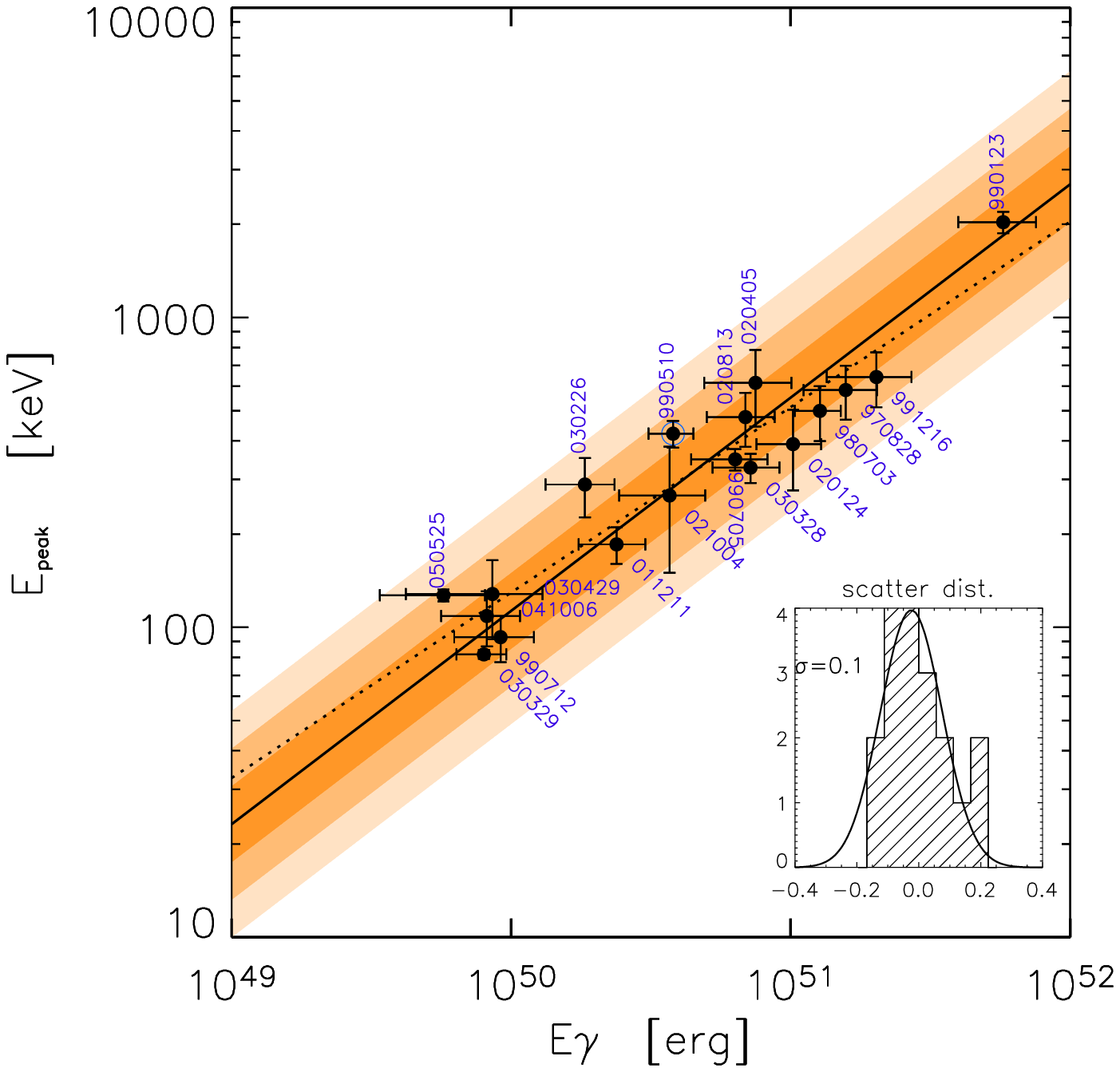}}
\caption{The $E_{\rm peak}$--$E_{\gamma}$ correlation from \cite{nava}
obtained under the assumption that the circum burst medium has a
constant density. The 18 bursts are those with measured redshift and
firm measurements of their spectral properties. The best powerlaw fit
of the data points, obtained by accounting for the uncertainties on
both coordinates, is shown (solid line) together with the linear
regression analysis, i.e. without accounting for the data
uncertainties (dotted line). The scatter of the data points around the
best fit correlation is shown in the insert with its Gaussian fit.
The 1,2 and 3$\sigma$ scatter of the data points around the best
correlation is represented by the shaded region. }\label{hm}
\end{center}
\end{figure}

\begin{figure}[t]
\begin{center}
\resizebox{12cm}{12cm}{\includegraphics{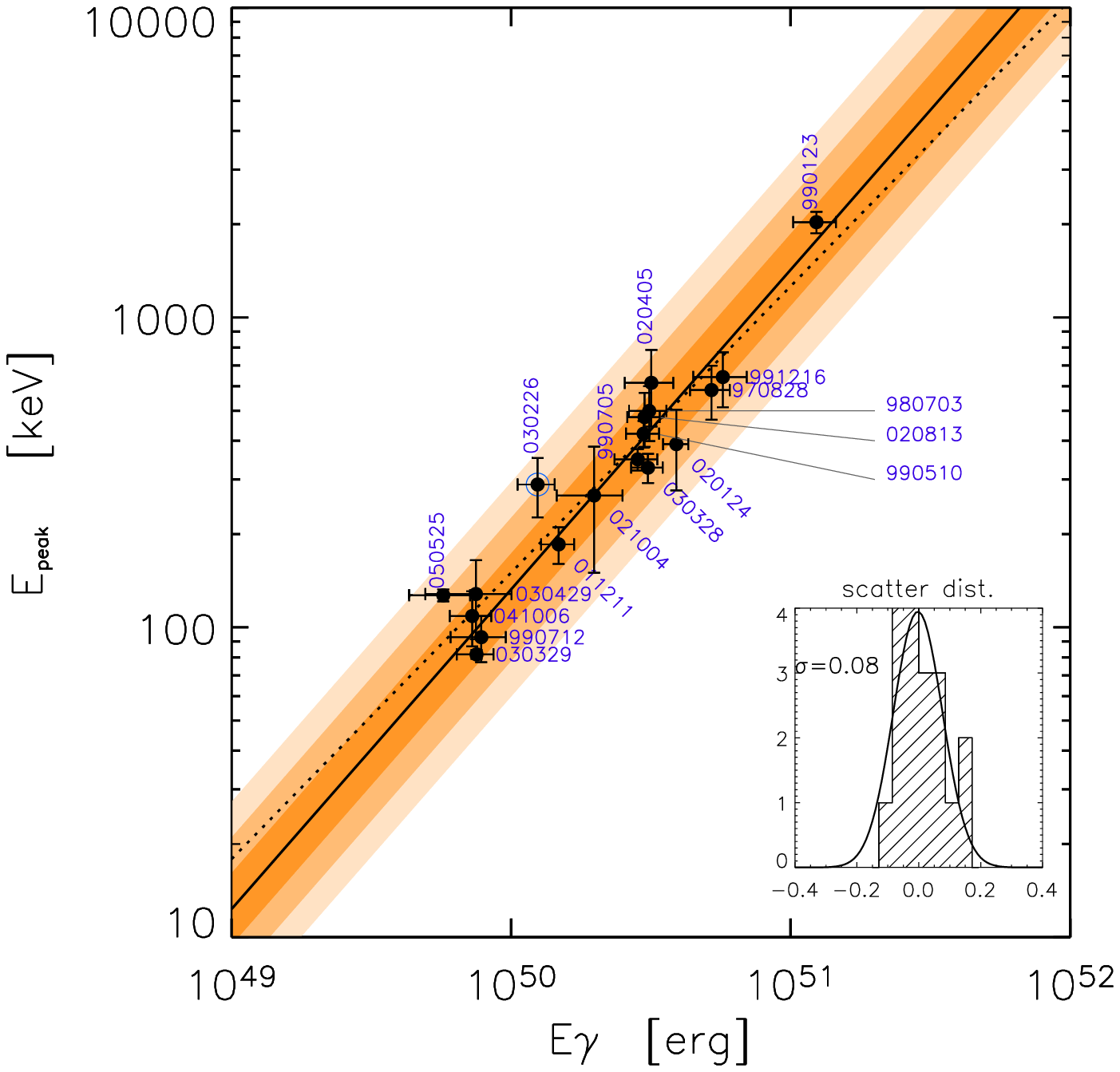}}
\caption{The $E_{\rm peak}$--$E_\gamma$ correlation in the case of a
wind profile of the external medium density as found with the 18 GRBs
reported in \cite{nava}.  The solid line represents the best fit
powerlaw model obtained accounting for the errors on both coordinates.
We also show the fit obtained with the simplest linear regression,
i.e. without accounting for the errors on the coordinates (dotted
line), which has a slope of 0.92.  The circled point represents GRB
030326 which is giving the largest contribution (23\%) to the best fit
reduced $\chi^2$.  The shaded regions represent the 1, 2 and 3$\sigma$
scatter around the best fit correlation.  The names of the 18 GRBs are
indicated.  The {\it insert} reports the distribution (hatched
histogram) of the scatter of the data points computed perpendicular
to the best correlation (solid line in the main plot) and its gaussian
fit (solid line in the insert) which has a $\sigma=0.08$. }
\label{wm}
\end{center}
\end{figure}

As shown above the correlations between the spectral peak energy and
the isotropic energy or luminosity are too much scattered to be used
as distance indicators. Only the $L_{\rm iso}$--$E_{\rm
peak}$--$T_{0.45}$, which is based on prompt emission properties, is
sufficiently tight to standardize GRB energetics (as it will be shown
in Sec.~5 and Sec.~6).

However, all the above correlations have been derived under the
hypothesis that GRBs are isotropic sources. Indeed, the possibility
that GRB fireballs are collimated was first proposed for GRB~970508
(\cite{waxman}) and subsequently invoked for GRB~990123 as a possible
explanation for their extraordinarily large isotropic energy
(\cite{fruchter}).  The main prediction of the collimated GRB model is
the appearance of an achromatic break in the afterglow light curve
which, after this break time, declines more steeply than before it
(\cite{rohads}, \cite{sari}).  Due to the relativistic beaming of the
photons emitted by the fireball, the observer perceives the photons
within a cone with aperture $\theta_{\Gamma}\propto 1/\Gamma$, where
$\Gamma$ is the bulk Lorentz factor of the material responsible for
the emission. During the afterglow phase the fireball is decelerated
by the circum burst medium and its bulk Lorentz factor decreases,
i.e. the beaming angle $\theta_{\Gamma}$ increases with time. A
critical time is reached when the beaming angle equals the jet opening
angle, i.e. $\theta_{\Gamma}\sim 1/\Gamma\sim\theta_{\rm jet}$,
i.e. when the entire jet surface is visible .  Under such hypothesis
the jet opening angle $\theta_{\rm jet}$ can be estimated
through this characteristic time (\cite{sari}), i.e. the so called
jet--break time $t_{\rm break}$ of the afterglow light curve.  Typical
$t_{\rm break}$ values ranges between 0.5 and 6 days
(\cite{frail},\cite{bloom},\cite{ghirlanda04})

\begin{table}\begin{center}
\caption{\label{tabella2}GRB intrinsic correlations involving collimation
corrected quantities. N: number of GRBs used to find the
correlation. The rank correlation coefficient $r_s$ and its chance
probability $P$ are given. The best fit parameters (with their 1$\sigma$
uncertainty) are also reported together with their $\chi^{2}$(dof).}
{\scriptsize
\begin{tabular}{@{}llllllll}
\br
Coll. corrected Correlations           &  N  & $r_{s}$ & $P$  &  $q$          &  $\alpha$      & $\beta$      & $\chi^{2}$(dof) \\
\hline
$Log(E_{\rm peak})= q + \alpha  Log(E_{\gamma,HM})$ & 18 & 0.9  & 2.3$\times10^{-8}$ &   -32.36$\pm$2.27  & 0.69$\pm$ 0.04   &    & 22.4(16)    \\
$Log(E_{\rm peak})= q + \alpha  Log(E_{\gamma,WM})$  & 18  & 0.92   & 6.8$\times10^{-8}$  & -49.44$\pm$3.27   & 1.03$\pm$0.06 &   & 18(16) \\
$Log(E_{\rm peak})= q + \alpha  Log(E_{\rm iso})$ & 18  &   &   & -48.05$\pm$0.24  & 1.93$\pm$0.11 & -1.08$\pm$0.17  & 24.2(16) \\
...... $+ \beta Log(t_{\rm break})$  &  &   &   &   &  &   &  \\
\hline
$Log(E_{\rm peak})= q + \alpha  Log(L_{\gamma,HM})$  &16 & 0.9  & 4$\times10^{-4}$   &  -22.8$\pm$1.3  & 0.51$\pm$0.03   &   &  41(14)\\
$Log(E_{\rm peak})= q + \alpha  Log(L_{\gamma,WM})$  &16 & 0.9  & 7$\times10^{-4}$   &  -35.1$\pm$2.0  & 0.75$\pm$0.04   &   &  76(14)\\ 
\br
\end{tabular}}\end{center}
\end{table}

The jet opening angle can be derived from $t_{\rm break}$ in two
different scenarios (e.g. \cite{chevalier}): (a) assuming that the
circum burst medium is homogeneous (HM) or (b) assuming a stratified
density profile (WM) produced, for instance, by the wind of the GRB
progenitor.  In both cases the jet is assumed to be uniform.

\section{The peak energy--collimation corrected energy correlation ($E_{\rm peak}$-$E_{\gamma}$)}

By correcting the isotropic energy of GRBs with measured redshifts,
under the hypothesis of a homogeneous circum--burst medium, Ghirlanda
et al. (\cite{ghirlanda04}) discovered that the collimation corrected
energy $E_{\gamma}=E_{\rm iso}(1-\cos\theta_{\rm jet})$ is tightly
correlated with the rest frame peak energy. With an initial sample of
15 GRBs (\cite{ghirlanda04}) this correlation results steeper (
i.e. $E_{\rm peak}\propto E_{\gamma}^{0.7}$) than the corresponding
correlation involving the isotropic energy and with a very small
scatter (i.e. $\sigma\sim 0.1$ to be compared with the $\sim 0.2$
scatter of the correlation with $E_{\rm iso}$). The addition of new
GRBs (some of which also discovered by {\it Swift} and, at least, by
another {\it IPN} satellite in order to have an estimate of the peak
energy) has confirmed this correlation and its small scatter
(e.g. \cite{ghirlanda05a}).

Recently, Nava et al. (\cite{nava}) reconsidered and updated the
original sample of GRBs with firm estimate of their redshift, spectral
properties and jet break times.  With the most updated sample of 18
GRBs we can re-compute the $E_{\rm peak}$--$E_{\gamma}$ correlation in
the HM case.  This is represented in Fig.\ref{hm} and results $E_{\rm
peak}\propto E_{\gamma}^{(0.69\pm0.04)}$ (see Tab.2) with a very small
scatter (i.e. $\sigma\sim0.1$).  Only two bursts (i.e. GRB~980425 and
GRB~031203 - which are not reported in the figures - but see
\cite{ghirlanda05a}) are outliers to this correlation, as well as to
the $E_{\rm peak}$-$E_{\rm iso}$ correlation. Possible interpretations
has been put forward such as the possibility that these events are
seen out of their jet opening angle (\cite{ramirez05}) and therefore
appear de-beamed in both $E_{\rm peak}$ and $E_{\rm iso}$ (see also
\cite{watson}). However, other possible explanations could be
considered (\cite{cobb},\cite{ghisellini06}).

Most intriguingly, \cite{nava} also computed the $E_{\rm
peak}$--$E_{\gamma}$ correlation in the WM case where a typical
$r^{-2}$ density profile is assumed. The resulting correlation is
reported in Fig.\ref{wm}. The $E_{\rm peak}$-$E_{\rm gamma}$
correlation derived in the WM case has two major properties: (i) its
scatter, i.e.  $\sigma=0.08$, is smaller than in the HM case
(although, due to the still limited number of data points, this does
not represent a test for the WM case against the HM case) and (ii) it
is linear, i.e. $E_{\rm peak}\propto E_{\gamma}$. Under the hypothesis
that our line of sight is within the GRB jet aperture angle, the
existence of a linear $E_{\rm peak}$-$E_{\gamma}$ correlation also
implies that it is invariant when transforming from the source rest
frame to the fireball comoving frame. A striking, still not explained,
consequence of this property is that the total number of photons
emitted in different GRBs is similar and should correspond to $\sim
10^{57}$ i.e. roughly the number of baryons in 1 solar mass. The
latter property might have important implications for the
understanding of the dynamics and radiative processes of GRBs.
 
\begin{figure}\begin{center}
\resizebox{12cm}{10cm}{\includegraphics{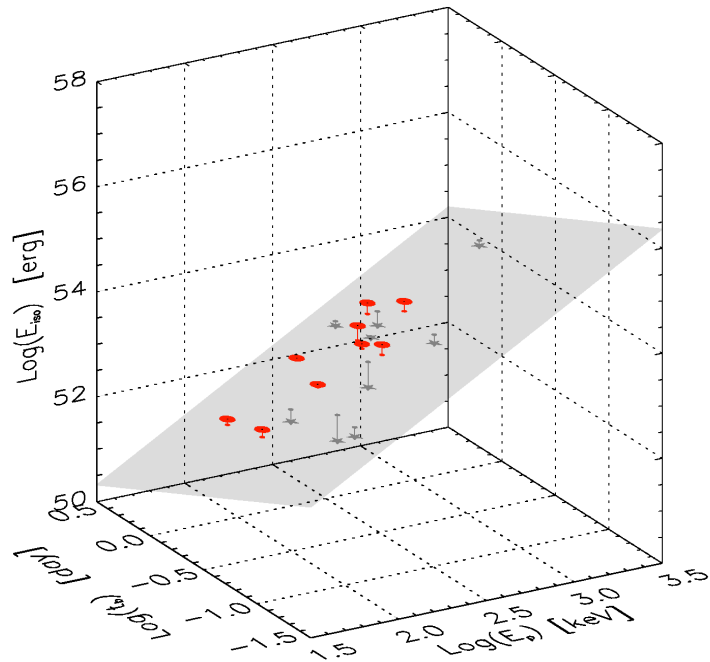}}
\caption{The 3-D representation of the $E_{\rm peak}$--$E_{\rm
iso}$--$t_{\rm break}$ correlation computed with the updated sample of
\cite{nava}. The shaded plane represents the best fit correlation and
the data points are marked differently if the lie above (red points)
or below (gray stars) this plane. The vertical lines show the distance
of the data points from the best fit plane. }
\label{zhang}
\end{center}
\end{figure}

One of the major criticism encountered by the above correlations
between the peak energy and the collimation corrected energy is the
fact that the collimation correction is derived from the measure of
$t_{\rm break}$ by assuming the standard fireball model. If on the one
hand the small scatter of the $E_{\rm peak}$-$E_{\gamma}$ correlation
might be, in turn, regarded as a confirm of this model, on the other
hand it could still be a matter of debate when these correlations are
used for cosmographic purposes.

\section{The peak energy--isotropic energy--jet break time correlation ($E_{\rm peak}$-$E_{\rm iso}$-$t_{\rm break}$)}

A completely empirical correlation also links the same three
observables used in deriving the above $E_{\rm peak}$-$E_{\gamma}$
correlation. Liang \& Zhang (\cite{liang}), in fact, found that the peak
energy $E_{\rm peak}$, the jet break time $t_{\rm break}$ and the
isotropic energy $E_{\rm iso}$ are correlated. In Fig.\ref{zhang}, we
report the 3-D correlation obtained with the most updated sample of 18
GRBs for which all the three observables are firmly measured. The
scatter of the data points around this correlation allows its use to
constrain the cosmological parameters (\cite{liang}). As discussed in
\cite{nava} the model dependent $E_{\rm peak}$-$E_{\gamma}$
correlations (i.e. derived under the assumption of a standard uniform
jet model and either for a uniform or a wind circumburst medium) are
consistent with this completely empirical 3-D correlation.  This
result, therefore, reinforces the validity of the scenario within
which they have been derived, i.e. a relativistically fireball with a
uniform jet geometry which is decelerated by the external medium, with
either a constant density or with an $r^{-2}$ profile.

Similarly to what has been done with the isotropic quantities, we can
explore if the collimation corrected $E_{\rm peak}$-$E_{\gamma}$
correlation still holds when the luminosity, instead of the energy, is
considered. In the second part of Tab.2 we report also the
correlations between $E_{\rm peak}$ and $L_{\gamma}$ in the HM case
(which represents an update of the correlation discussed in
\cite{ghirlanda05b}) and also the new correlation in the WM case. We
note that both the HM and the WM correlations defined with
$L_{\gamma}$ are more scattered than the corresponding correlations
defined with the collimation corrected energy $E_{\gamma}$.

\subsection{Probing the $E_{\rm peak}-E_{\rm iso}$ correlation}

One of the major drawback of the $E_{\rm peak}-E_{\rm iso}$
correlation is that it has been found with a small number of
GRBs. This led to think that some selection effects might play a
relevant role in deriving this correlation. In support of this
correlation we should mention that the extension
(\cite{ghirlanda04},\cite{amati06}) of the original sample of GRBs
from which it was derived (\cite{amati}) leads to find again a
statistically significant correlation between $E_{\rm peak}$ and
$E_{\rm iso}$. 
\begin{figure}\begin{center}
\resizebox{13cm}{11cm}{\includegraphics{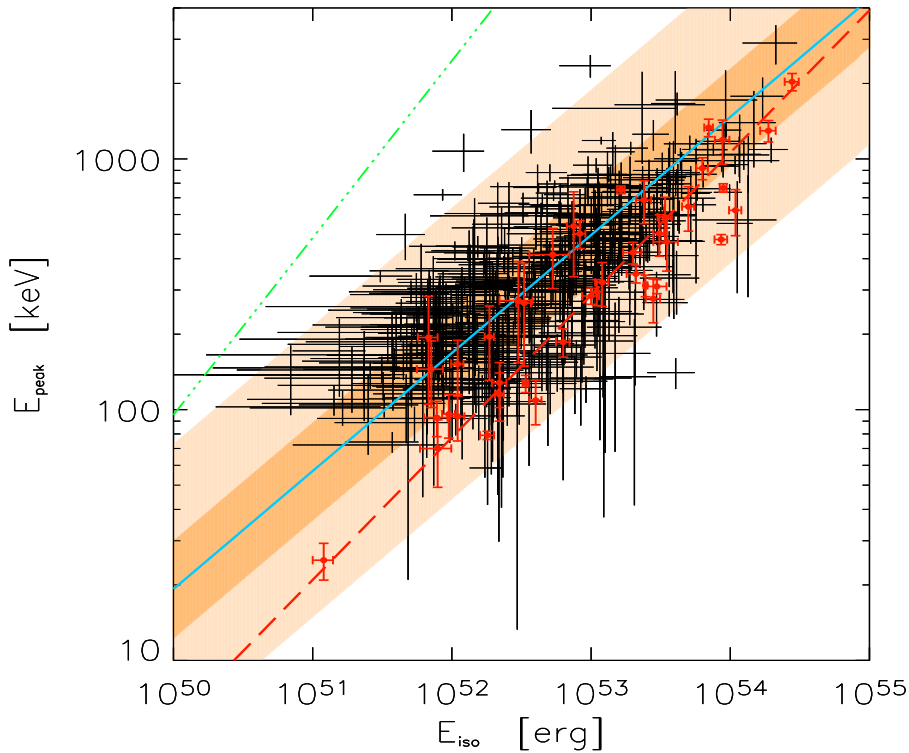}}
\caption{Rest frame peak energy $E_{\rm peak}$ versus isotropic
equivalent energy $E_{\rm iso}$. Red filled circles are the 43
GRBs with spectroscopically measured redshifts and published spectral
properties (adapted from \cite{amati06}). The long--dashed line is
their best fit (weighting for the errors on both variables) which is
reported in Tab.~1. The black crosses are the 442 GRBs with pseudo
redshifts derived from the lag--luminosity relation. The solid line is
the best fit to these data points which results $E_{\rm peak}\propto
E_{\rm iso}^{0.47}$ with a reduced $\chi^2_{\rm r}=4.0$ (440
dof). The shaded region represents the 3$\sigma$ scatter of the black
points around their best fit line (solid line). The
triple--dot--dashed line represents the $E_{\rm peak}-E_{\gamma}$
correlation.}
\label{probing}\end{center}
\end{figure}

One test that can be performed even without knowing the redshift of
GRBs is to check if bursts with measured fluence $S$ and peak energy
$E_{\rm peak}^{\rm obs}$ are consistent with the $E_{\rm peak}-E_{\rm
iso}$ correlation, by considering all possible redshifts. The
application of this test to a sample of BATSE bursts (\cite{nakar},
\cite{band05}) led to claim that the $E_{\rm peak}$--$E_{\rm iso}$
correlation is not satisfied by most of the BATSE GRBs without a
redshift measurement. Instead, a different conclusion has been reached
by \cite{bosnjak}.

We performed (\cite{ghirlanda05b}) a different test by checking if a
large sample of GRBs with a pseudo-redshift measurement could indicate
the existence of a relation in the $E_{\rm peak}-E_{\rm iso}$
plane, although with a possible larger scatter than that found for the
$E_{\rm peak}$-$E_{\rm iso}$ correlation with few GRBs with
spectroscopic measured redshifts.

\begin{figure}\begin{center}
\resizebox{12cm}{10cm}{\includegraphics{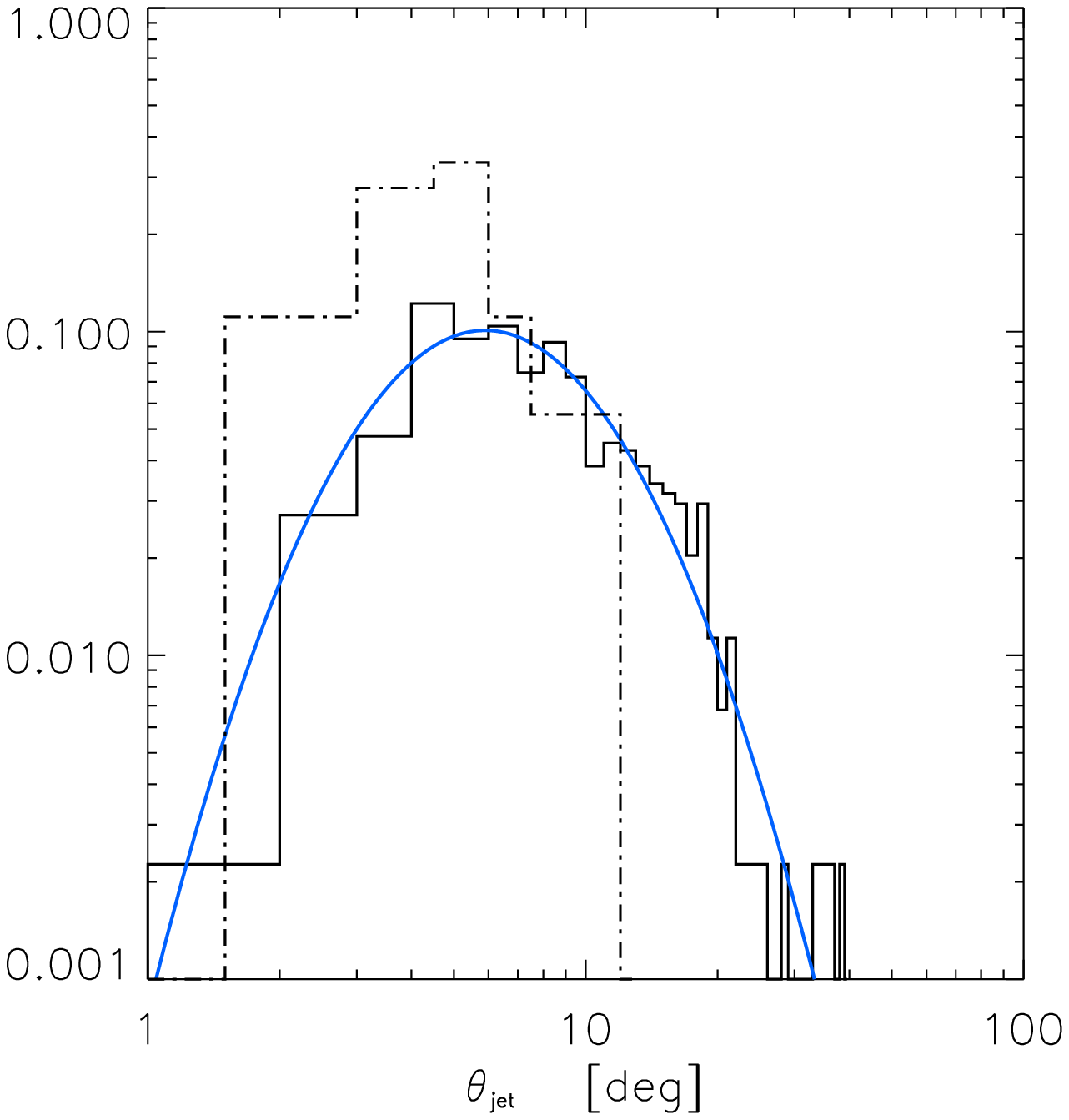}}
\caption{Jet opening angle distributions. The solid histogram
  represents the $\theta_{\rm jet}$ derived from the large sample of
  442 GRBs with pseudo redshifts requiring that they satisfy the
  $E_{\rm peak}-E_{\gamma}$ correlation
  (\cite{ghirlanda05a},\cite{nava} - with the approximation of no
  scatter). The solid line is the best fit log--normal
  distribution. The dot-dashed histogram represents the angle
  distribution of the 19 GRBs with spectroscopic redshifts and well
  constrained $t_{\rm jet}$.}
\label{angoli}\end{center}
\end{figure}

With a sample of 442 long duration GRBs whose spectral properties has
been studied (\cite{yonetoku}) and whose redshifts has been derived
from the spectral lag--luminosity correlation (\cite{band04}), we
populated the $E_{\rm peak}-E_{\gamma,\rm iso}$ plane (black crosses
in Fig.~\ref{probing}).  We found that this large sample of burst
produce an $E_{\rm peak}$-$E_{\rm iso}$ correlation (solid line in
Fig.\ref{probing}) whose slope (normalization) is slightly flatter
(smaller) than that found with the sample of 43 GRBs (red filled
circles in Fig.\ref{probing}). The gaussian scatter of the
pseudo--redshift bursts around their correlation has a standard
deviation $\sigma=0.2$, i.e. consistent with that of the 43 GRBs
around their best fit correlation (long-dashed red line in
Fig.\ref{probing}). This suggests that indeed a correlation between
$E_{\rm peak}$ and $E_{\rm iso}$ exists.

However, it might still be argued that the correlation found with the
43 and with the 442 GRBs have inconsistent normalization, although
similar scatter and slopes. This is shown by the 43 GRBs (red filled
circles in Fig.\ref{probing}) with spectroscopically measured
redshifts which lie on the right tail of the scatter distribution of
the 442 GRBs (black crosses) in the $E_{\rm peak}-E_{\rm iso}$
plane. This fact might be interpreted as due to the selection effect
of the jet opening angle. In fact, from the scatter of the 442 GRBs we
can derive their jet opening angles by assuming that the $E_{\rm
peak}$-$E_{\gamma}$ correlation exists and that its scatter is (as
shown in \cite{ghirlanda04}) much smaller than that of the $E_{\rm
peak}-E_{\rm iso}$ correlation. The $\theta_{\rm jet}$
distribution derived in this way is shown in Fig.\ref{angoli} (solid
histogram) and it is well represented by a log-normal distribution
(solid line) with a typical $\theta_{\rm jet}\sim 6^{o}$. The
$\theta_{\rm jet}$ distribution of the 19 GRBs with measured $t_{\rm
jet}$ is also shown (dot-dashed histogram in Fig.\ref{angoli}) and it
appears shifted to the small--angle tail of the $\theta_{\rm jet}$
distribution of the 442 angles. This suggests that the 43 GRBs, which
are used to define the $E_{\rm peak}-E_{\rm iso}$ correlation, have
jet opening angles which are systematically smaller than the average
value of the distribution and this is what makes them more luminous
and, therefore, more easily detected.

Interestingly the fact that the scatter of the 442 GRBs in the $E_{\rm
peak}-E_{\rm iso}$ plane around their best fit correlation is gaussian
indicates that, if the scatter is due solely to their angle
distribution, this is characterized by a preferential angle. If,
instead, the angle distribution were flat, we should expect a
uniform distribution of the data points around the $E_{\rm
peak}-E_{\rm iso}$ correlation.

\begin{figure}[htb]
\begin{minipage}[t]{8cm}
\resizebox{8cm}{7cm}{\includegraphics{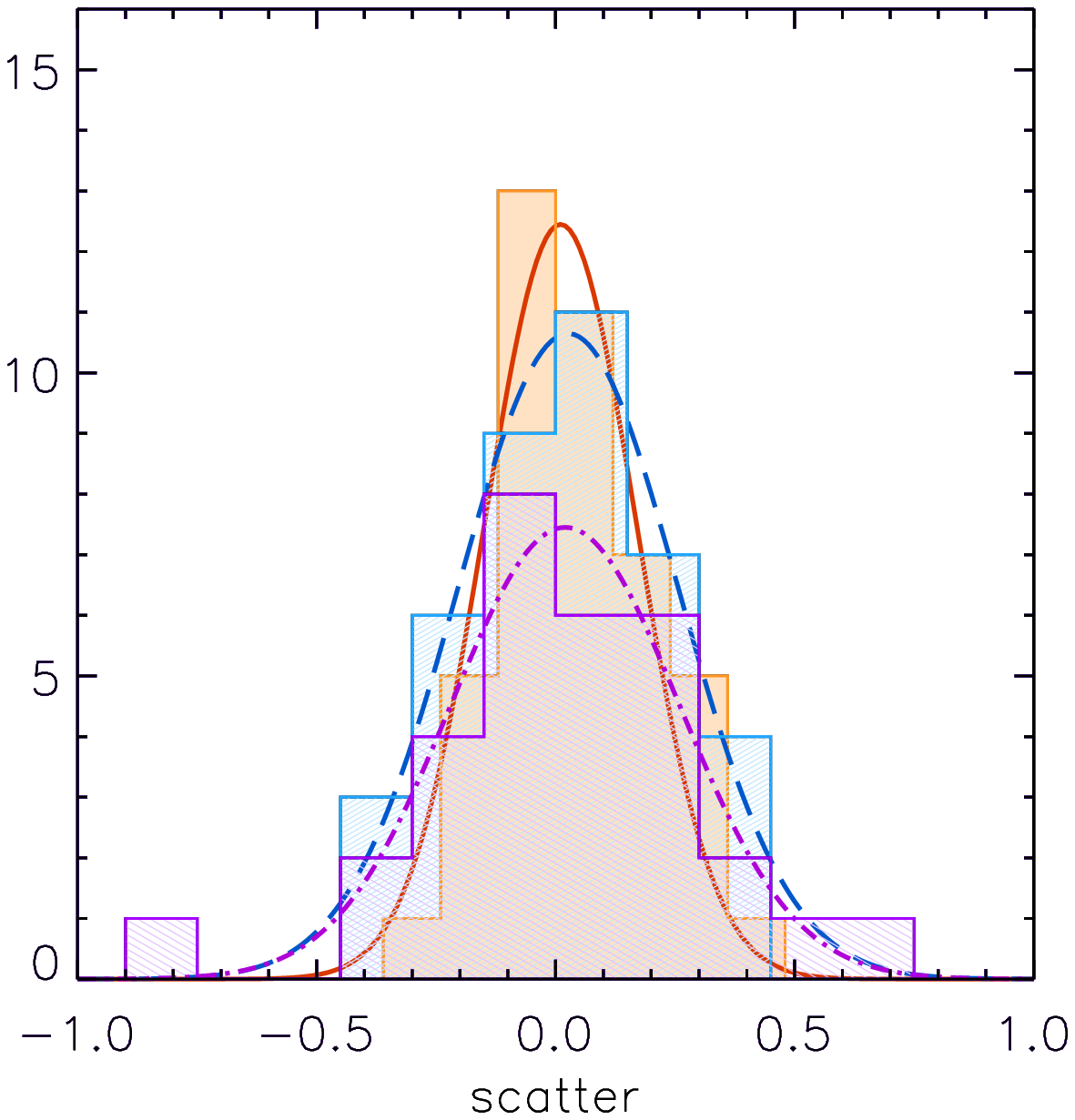}}
\end{minipage}\hfill
\begin{minipage}[t]{8cm}
\resizebox{8cm}{7cm}{\includegraphics{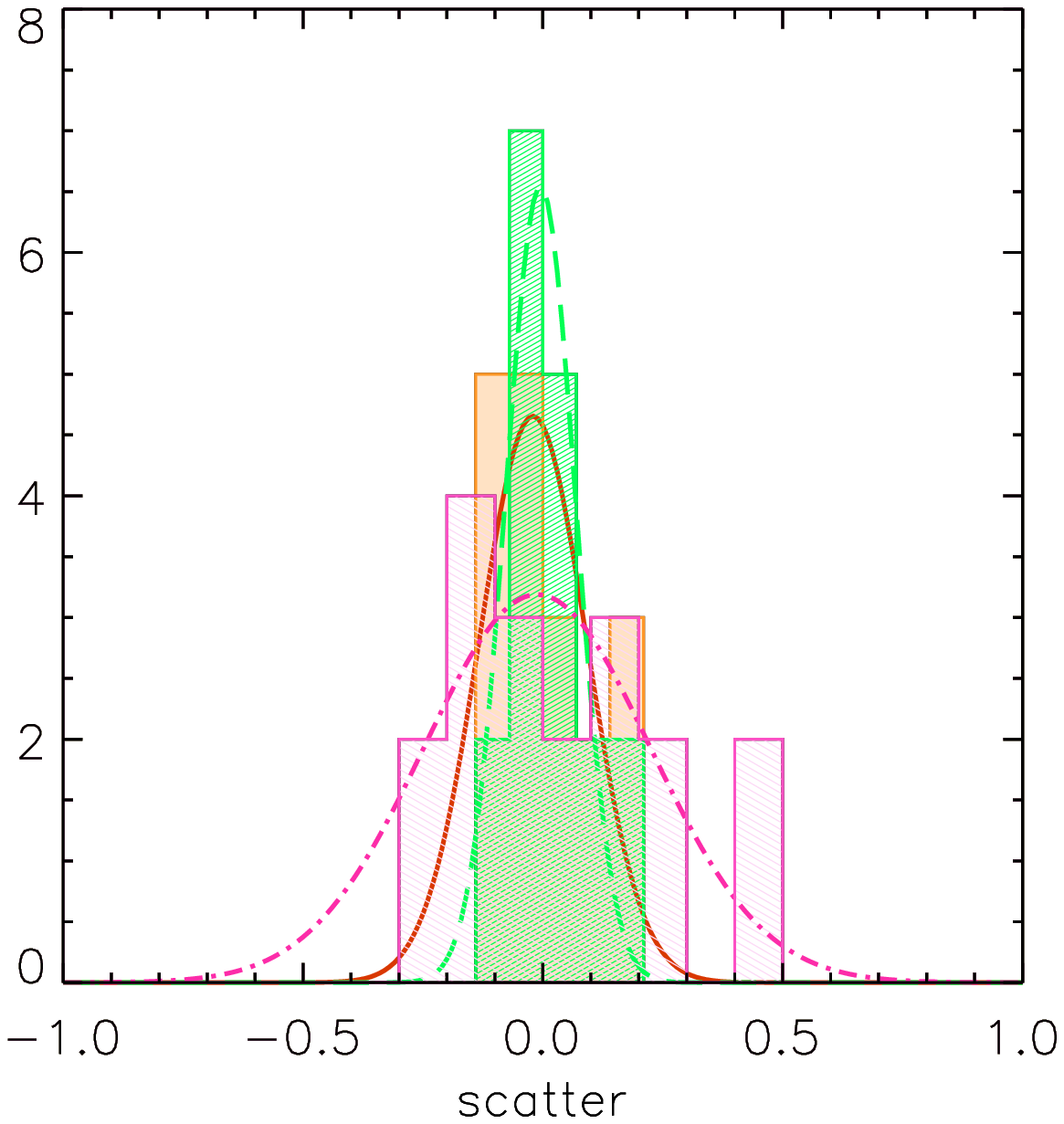}}
\end{minipage}
\caption[]{{\bf Left}: Distribution of the scatter around the
correlations involving isotropic quantities (reported in Tab.1).  The
orange distribution (and solid orange line) represents the scatter
distribution (and its gaussian fit) around the $E_{\rm peak}$-$E_{\rm
iso}$ correlation. The blue histogram and violet histograms are for
the $E_{\rm peak}$-$L_{\rm iso}$ and $V-L_{\rm iso}$ correlation,
respectively. {\bf Right}: Distribution of the scatter around the
correlations involving the collimation corrected quantities and of the
3-D empirical correlation (reported in Tab.2). The green distribution
(and dashed green line) represents the scatter distribution (and its
gaussian fit) around the $E_{\rm peak}$-$E_{\gamma}$ correlation (in
the wind circumburst scenario). The orange and pink histograms are for
the scatter of the $E_{\rm peak}$-$E_{\gamma}$ (in the homogeneous
circumburst medium scenario) and $E_{\rm iso}-E_{\rm peak}-t_{\rm
break}$ correlation, respectively.}
\label{scat}
\end{figure}

\section{Constraining the cosmological parameters with GRBs}

The scatter of the correlations described in the previous sections are
compared in Fig.\ref{scat}. The collimation corrected correlations
$E_{\rm peak}$--$E_{\gamma,HM}$ and $E_{\rm peak}$--$E_{\gamma,WM}$ or
the empirical correlation $E_{\rm peak}$--$E_{\rm iso}$--$t_{\rm
break}$, due to their very low scatter can be used to constrain the
cosmological parameters. Moreover, all these correlations have an
acceptable reduced $\chi^2$ in the concordance cosmological model.
 
\begin{figure}\begin{center}
\resizebox{12cm}{10cm}{\includegraphics{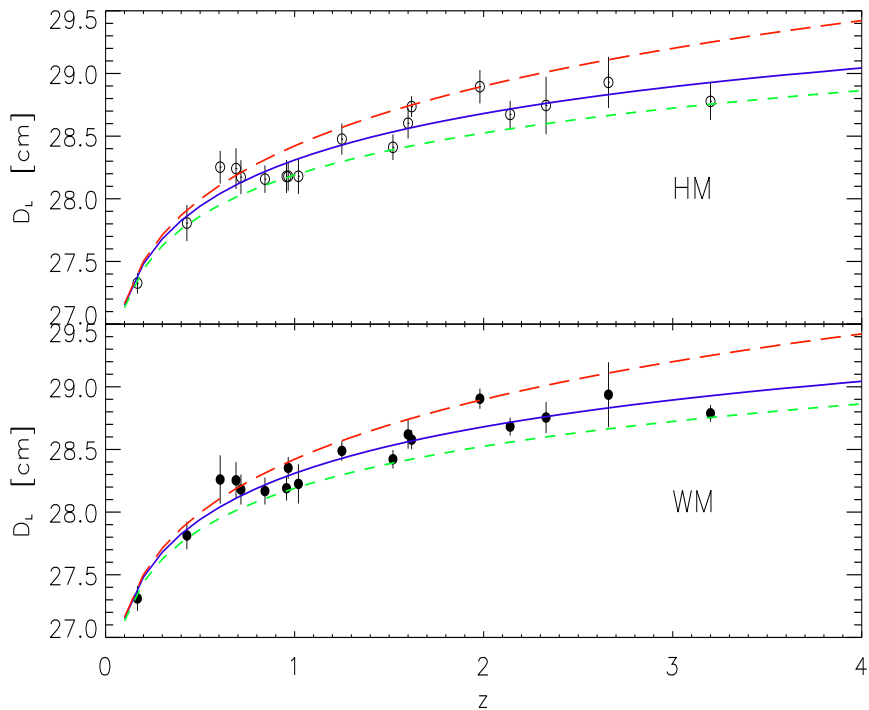}}
\caption{Hubble diagram of the 18 GRBs of the sample of \cite{nava}
derived with the $E_{\rm peak}$--$E_{\gamma}$ correlations in the HM
and WM case. The curves represent different cosmological models:
($\Omega_{M},\Omega_\Lambda$)=(0.3,0.7) [solid--blue line],
($\Omega_{M},\Omega_\Lambda$)=(0.0,1.0) [long--dashed red line],
($\Omega_{M},\Omega_\Lambda$)=(1.0,0.0) [short--dashed green line].}
\label{hd_corr}
\end{center}
\end{figure}

As a first test of the possible use of the above correlations for
cosmography we show in Fig.~\ref{hd_corr} the luminosity distance
derived with the $E_{\rm peak}$--$E_{\gamma}$ correlation in the HM
and WM scenario for the 18 GRBs of the sample discussed in
\cite{nava}. By comparing the plots with Fig.~\ref{hd} we note that by
using the $E_{\rm peak}$-$E_{\gamma}$ correlation the dispersion of
the data points is reduced with respect to the assumption that
all GRBs have a unique luminosity (see Sec.~2). Also note that no
substantial difference appears if using the $E_{\rm
peak}$-$E_{\gamma}$ correlation derived either in the HM or in the WM
case (top and lower panel of Fig.~\ref{hd_corr} respectively).

Different methods have been proposed to constrain the cosmological
parameters with GRBs (\cite{ghirlanda04a}, \cite{dai},
\cite{firmani}). A common feature of these methods is that they
construct a $\chi^2$ statistic, based on one of the correlations
described before, as a function of two cosmological parameters. The
minimum $\chi^2$ and the statistical confidence intervals (i.e. 68\%,
90\% and 99\%) represent the constraints on the couple of cosmological
parameters.

One important point (see also Sec~7.3) is that the correlations that
we have presented so far have been derived assuming a particular
cosmological model, i.e. $\Omega_{M}$=0.3, $\Omega_{\Lambda}$=0.7 and
$h$=0.7. Indeed, in the $E_{\rm peak}$--$E_{\gamma,HM}$ correlation,
for instance, $E_{\rm peak}=E_{\rm peak}^{\rm obs}(1+z)$ is
independent from the cosmological model  whereas $E_{\gamma}$
depends on the cosmological parameters through the luminosity
distance, i.e. $E_{\gamma}=f(4\pi D_{L}^{2} F)/(1+z)$, where $D_{L}$
is a function of ($\Omega_{M}$,$\Omega_\Lambda$,$h$) (see
Sec.1.1). The need to assume a set of cosmological parameters to
compute the above correlations is mainly due to the lack of a set of
low redshift GRBs which, being cosmology independent, would calibrate
these correlations. Therefore, if we want to fit any set of cosmological
parameters with the above correlation we cannot use a {\it unique}
correlation i.e. obtained in a particular cosmology. This method,
which has indeed been adopted by \cite{dai}, would clearly suffer of
such a circular argument.

The ``circularity problem'' (\cite{ghisellini}) can be avoided in two
ways: (i) through the calibration of these correlations by several low
redshift GRBs (in fact, at $z\leq 0.1$ the luminosity distance has a
negligible dependence from any choice of the cosmological parameters
$\Omega_{M}$,$\Omega_\Lambda$) or (ii) through a solid physical
interpretation of these correlation which would fix their slope
independently from cosmology. A third possibility, presented in
Sec.~7, consists in calibrating the correlations with a set of GRBs
with a low dispersion in redshift.

While waiting for a sample of calibrators or for a definitive
interpretation, we can search for suitable statistical methods to use
these correlations as distance indicators but avoiding the
``circularity problem''. The methods that have been used to fit the
cosmological parameters through GRBs can be summarized as follows:
\begin{itemize}
\item (I) The {\it scatter method} (\cite{ghirlanda04a}). By fitting
the correlation for every choice of the cosmological parameters that
we want to constrain (e.g. $\Omega_{M},\Omega_{\Lambda}$) a $\chi^2$
surface (as a function of these parameters) is built. The best
cosmological model corresponds to the minimum scatter around the
correlation (which has to be re-computed in every cosmology in order
to avoid the circularity problem) and is identified by the minimum of
the $\chi^{2}(\Omega_{M},\Omega_{\Lambda})$ surface.
\item (II) The {\it luminosity distance method}
(\cite{ghirlanda04a},\cite{dai},\cite{liang}).  The main steps are:
(1) choose a cosmology and fit the $E_{\rm peak}$--$E_\gamma$
correlation ; (2) from the best fit correlation the term $E_{\gamma}$
is estimated; (3) from the definition of $E_{\gamma}=f\cdot E_{\rm iso}$
(where also $f=1-\cos\theta_{\rm jet}$ is a function of $E_{\rm iso}$
through $\theta_{\rm jet}$) derive $E_{\rm iso}$ from which (4) the
luminosity distance $D_{L,c}$ is computed; (5) build a $\chi^2$ by
comparing $D_{L,c}$ with that derived from the cosmological model
$D_{L}$. By repeating these steps for every choice of the cosmological
parameters a $\chi^{2}(\Omega_{M},\Omega_{\Lambda})$ surface is
derived. Also in this case the minimum $\chi^2$ represents the best
cosmology, i.e. the model which best matches the luminosity distance
derived from the correlation and that predicted by the model itself.
\item (III) The {\it Bayesian method} (\cite{firmani}). The two
methods described above are based on the concept that a correlation
exists between two variables, e.g. $E_{\rm peak}$ and
$E_{\gamma}$. However these correlations are also very likely related
to the physics of GRBs and, therefore, they should be unique. This
property is not exploited by methods (I) and (II). Firmani et
al. (\cite{firmani}) proposed a more sophisticated method which makes
use of both these conditions, i.e. the existence and uniqueness of the
correlation. The method is based on the Bayes theorem
(e.g. \cite{wall}). The basic steps are: (1) choose a cosmology
$\bar{\Omega}$ and fit the correlation; (2) test this correlation
(i.e. keep it fixed) in all the possible cosmologies $\Omega$ and
derive a conditioned probability surface $P(\Omega | \bar{\Omega})$
which represents a ``weight'' for the starting cosmology
$\bar{\Omega}$ (i.e. it describes how all the cosmologies that have
been tested ``respond'' to the starting cosmology); (4) by repeating
these steps for different starting cosmologies $\bar{\Omega}$ and
combining, at each step, the new conditioned probability with those
found in the previous iterations, the Bayesian methods naturally
converges to a final probability surface whose maximum represents the
best cosmological model.
\end{itemize}

\begin{figure}\begin{center}
\resizebox{13cm}{12cm}{\includegraphics{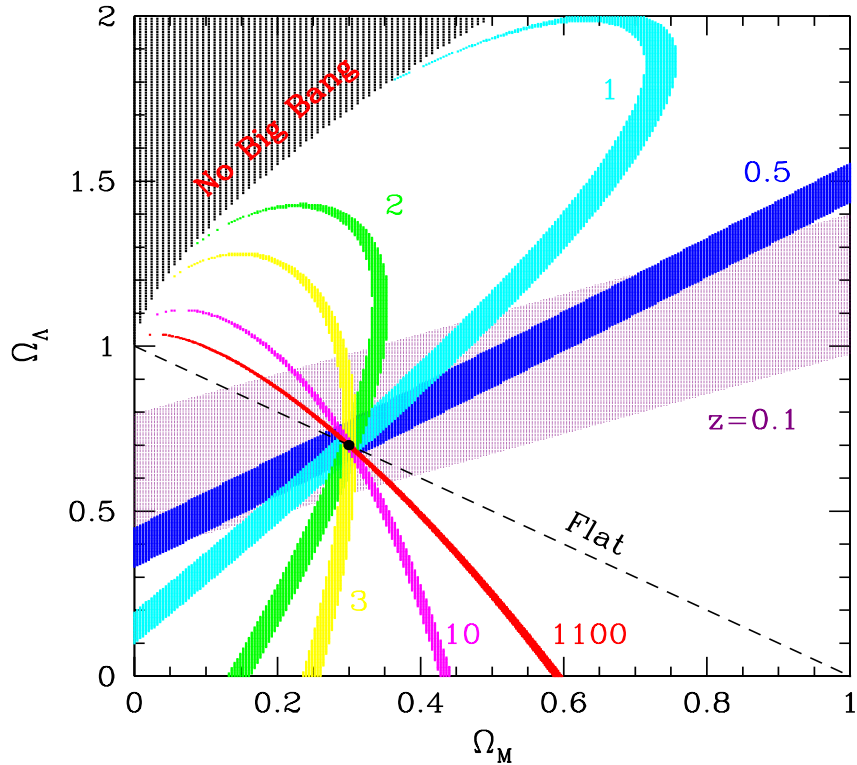}}
\caption{Regions of $\pm$1\% variation around lines of constant
$d_{\rm L}$ in the ($\Omega_{\rm M}$,$\Omega_{\Lambda}$) plane at
redshift 0.5,1,1.5 and 2. assuming that each line passes through the
fiducial point($\Omega_{\rm M}$=0.33,$\Omega_{\Lambda}$=0.77).
}\label{sbisciole}\end{center}
\end{figure}

The last method, which exploits all the information contained in the
adopted correlation, has also a practical advantage: it solves the
problem of the loitering line which separates Big-Bang and no-Big-Bang
universes. As shown in Fig.~\ref{sbisciole}, near this line the
luminosity distance stripes are highly wounded--up and clustered. In
correspondence to this line in fact $D_{L}$ falls rapidly to zero. If
the adopted sample of standard candles extends to very high redshifts,
this region can attract the minimum of the $\chi^2$ surface, as shown
in Fig.1 of \cite{ghirlanda04a}. For this reason the Bayesian method
is more accurate because it assigns a low probability to the
cosmologies near this line (see \cite{firmani}).

\section{Constraints on $\Omega_{M}$ and $\Omega_{\Lambda}$}

\begin{figure}\begin{center}
\resizebox{13cm}{12cm}{\includegraphics{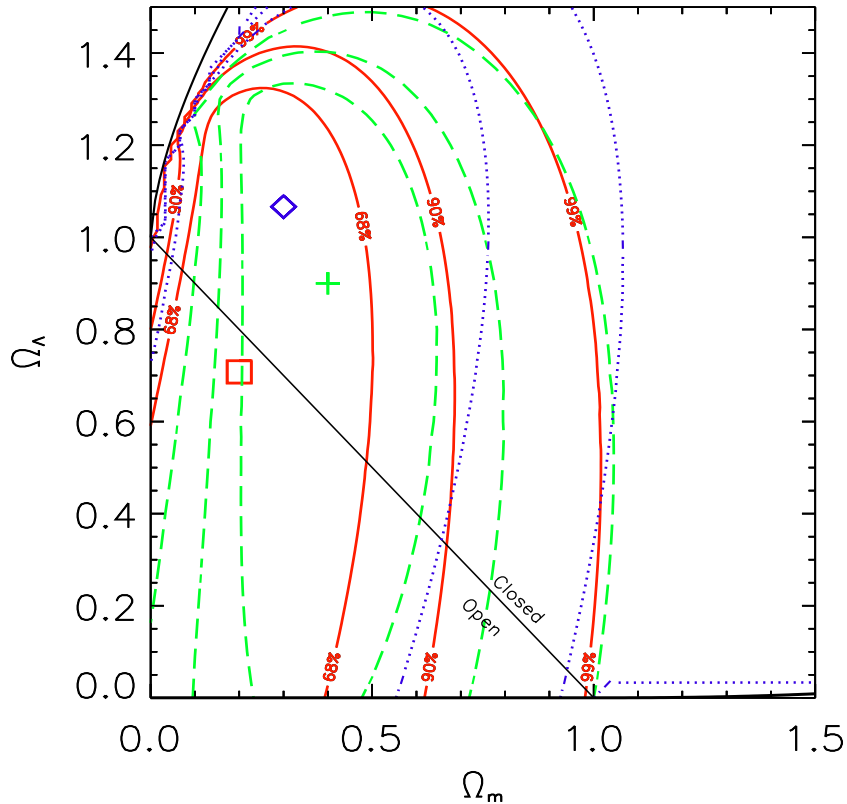}}
\caption{Constraints on the $\Omega_{M}$ and $\Omega_{\Lambda}$
parameters obtained with the Bayesian method through the three
correlations: $E_{\rm peak}$--$E_{\gamma,HM}$ (solid red line),
$E_{\rm peak}$--$E_{\gamma,WM}$ (dashed green line) and the empirical
correlation $E_{\rm peak}$--$E_{\rm iso}$--$t_{\rm break}^\prime$
(dotted blue line). The lines correspond to the 68\%, 90\% and 99\%
confidence contours.}\label{conf}\end{center}
\end{figure}

With the Bayesian method we can constrain the cosmological parameters
$\Omega_{M}$ and $\Omega_{\Lambda}$ with the sample of 18 GRBs
reported in \cite{nava} updated with GRB~051022 (see
\cite{ghirlanda06}). In Fig.~\ref{conf} we show the contours
corresponding to the 68\%, 90\% and 99\% confidence levels obtained
with all the three correlations, i.e. $E_{\rm peak}$--$E_{\gamma,HM}$
(solid line), $E_{\rm peak}$--$E_{\gamma,WM}$ (dashed line) and the
empirical 3D correlation $E_{\rm peak}$--$E_{\rm iso}$--$t_{\rm
break}$ (dotted line). We note that the contours obtained with these
correlations (of which the first two are model dependent and the third
is completely empirical) are consistent one with another.  We also
note that the centroid of the confidence contours are only slightly
different. However, given the large allowance area of the same
contours the exact position of the centroid is not relevant, i.e. the
contours obtained with GRBs as standard candles are consistent with
the concordance model.

\subsection{Statistical errors}

One of the critical issue in the use of the three correlations for
cosmology is related to statistical errors. There are, in fact, two
major sources of errors which shape the contours obtained with GRBs:
the statistical errors associated with the observables used to
calculate $E_{\rm peak}$ and $E_{\gamma}$ and the errors on the best
fit parameters (slope and normalization) of the correlation resulting
from the fitting procedure. These errors enters in the construction of
the $\chi^2$ surface.  

\begin{figure}[htb]
\begin{minipage}[t]{8cm}
\resizebox{8cm}{7cm}{\includegraphics{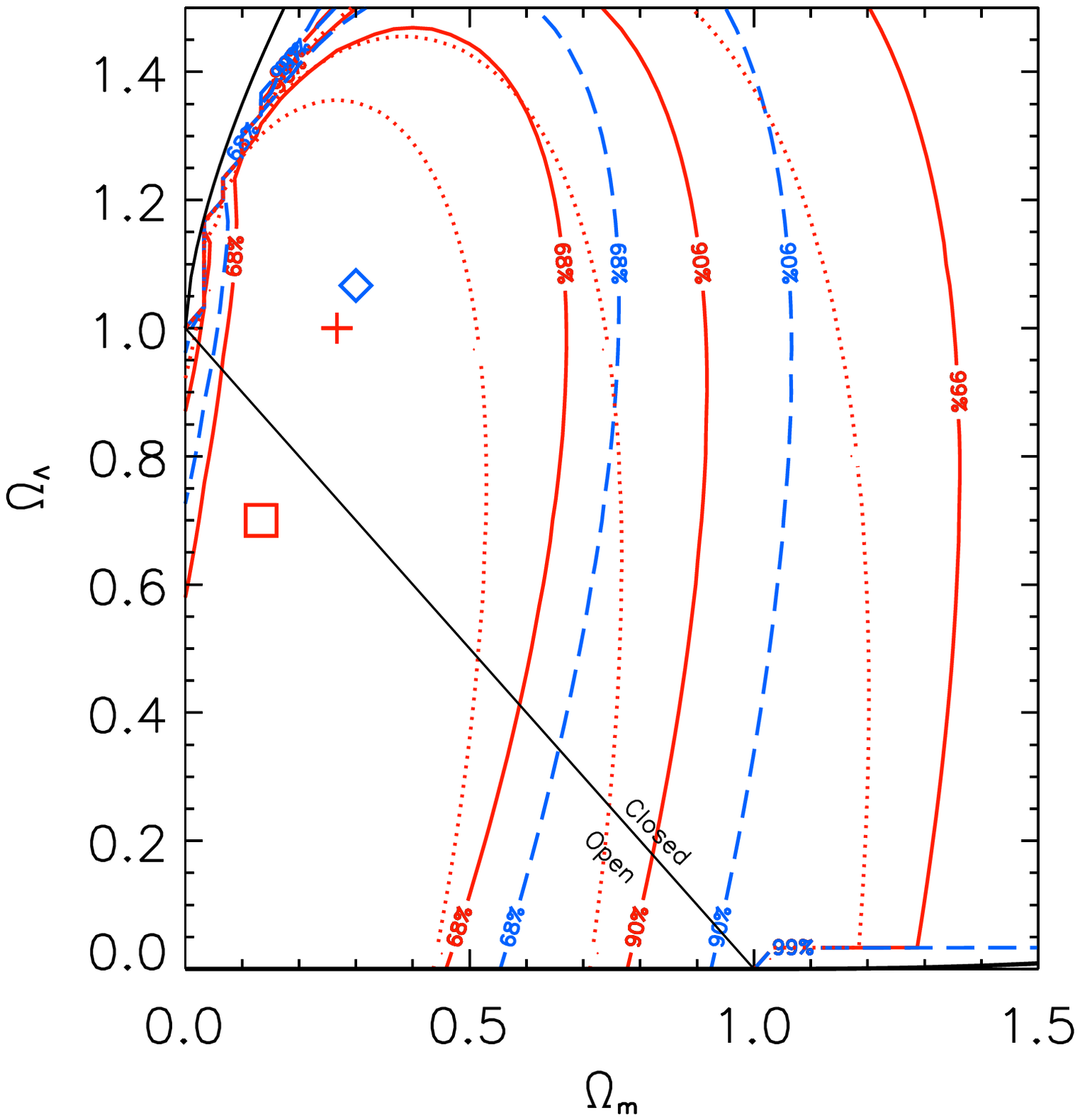}}
\end{minipage}\hfill
\begin{minipage}[t]{8cm}
\resizebox{8cm}{7cm}{\includegraphics{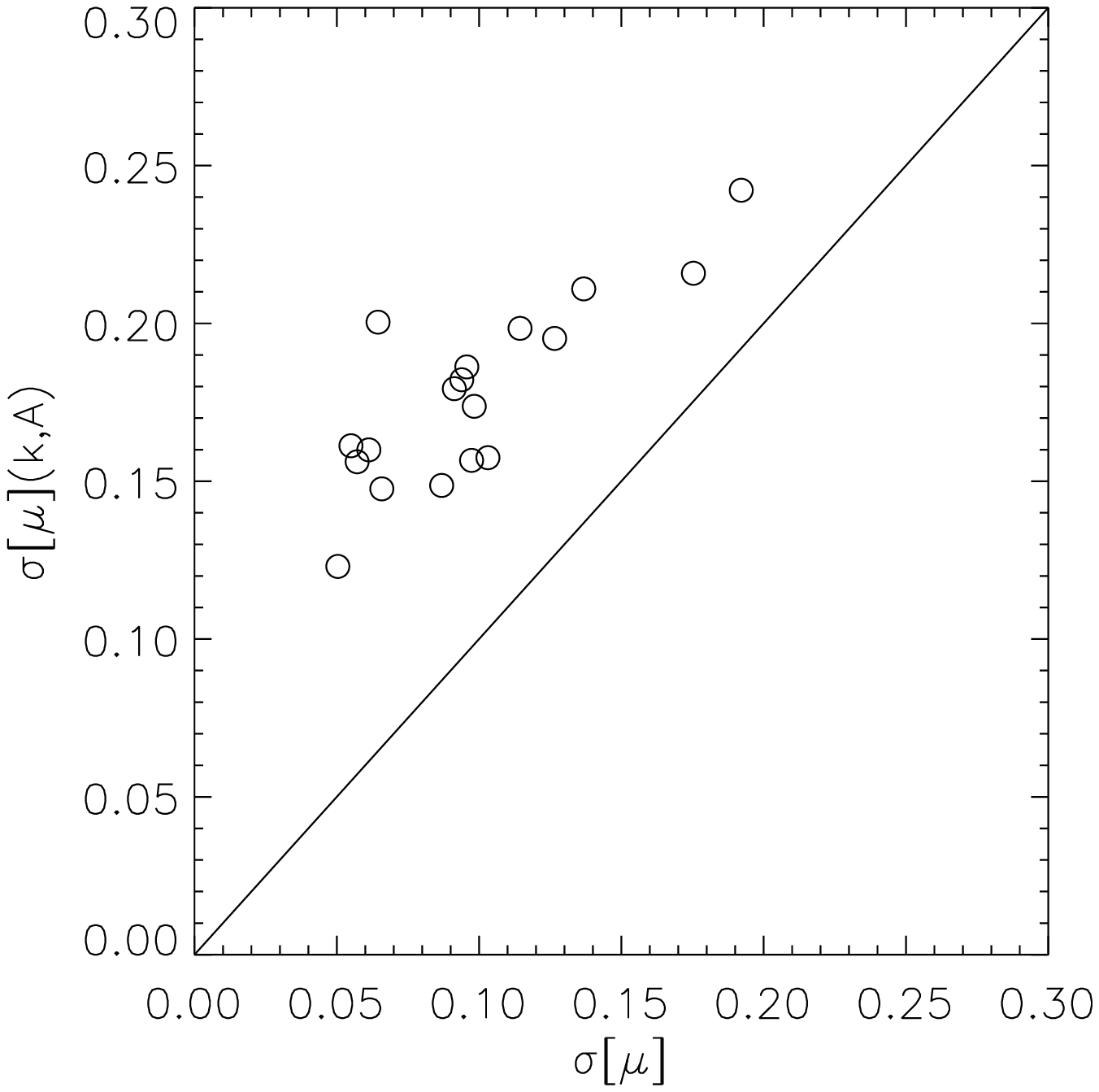}}
\end{minipage}
\caption[]{{\bf Left}:Constraints on the cosmological parameters
$\Omega_{M}$,$\Omega_{\Lambda}$ obtained with the empirical 3D
correlation and the updated sample of 19 GRBs. The solid (red)
contours are obtained by considering, as in \cite{liang}, only the
errors on the variables, i.e.  $E_{\rm peak}$, $E_{\rm iso}$ and
$t_{\rm break}$. The long--dashed blue contours include also the
errors on the best fit parameters. The dotted contours are obtained
with the original sample of 14 GRBs used by \cite{liang}, accounting
only for the errors on the variables. {\bf Right}: Errors on the
distance modulus $\mu$ computed considering the errors on the best fit
correlation parameters (k,A) versus the same error computed ignoring
these uncertainties and considering only the errors on $E_{\rm \gamma,
iso}$, $E_{\rm p}$ and $t_{\rm break}$. In this case we assumed that the
exponent of $t_{\rm break}$ is linear and we ignored the error on this
parameter.  }
\label{error}
\end{figure}

In particular, the issue of accounting for all the sources of errors
makes the contours particularly large as in the case of the empirical
3D correlation which involves three observables (with their
statistical uncertainties) and three fit parameters (with their
errors, resulting from the best fit of the data points).  As a simple
test we compare in Fig.~\ref{error} the constraints on $\Omega_{M}$
and $\Omega_{\Lambda}$ when accounting only for the errors on the
observables (solid--red contours) or when accounting also for the
errors on the best fit parameters (dashed blue contours). In
Fig.~\ref{error} (left panel) we show the constraints obtained with
the empirical $E_{\rm peak}$--$E_{\rm iso}$--$t_{\rm break}$
correlation by ignoring (solid contours) the errors associated to the
the best fit parameters which in this case are the normalization and 2
slopes (because the correlation is expressed, for instance, as $E_{\rm
iso}$ as a function of $E_{\rm peak}$ and $t_{\rm break}$). The same
figure shows, instead, that if we account for all the errors, we find
(dashed blue contours) considerably larger constraints. This
difference is also represented in the right panel of Fig.~\ref{error}
where the error on the distance modulus $\mu$ obtained accounting for
all the errors is represented versus the same quantity obtained
including only the errors on the observables.

The dotted (red) contours in Fig.~\ref{error} (left panel) are
obtained with the sample of 14 GRBs reported in the original work of
\cite{liang} without including the errors on the best fit
parameters. Here we adopted the Bayesian method to derive all the
cosmological contours. Moreover, another possible reason for the
difference between our contours (red dotted in Fig.~\ref{error}) and
the slightly smaller contours reported in \cite{liang} might be the
fact that our method for fitting the 3D empirical correlation takes
into account the errors on the three variables and adopts a the
$\chi^2$ minimization method to find the best fit parameter values,
whereas the multiregression analysis (\cite{liang}) does not account
for all the errors on the observables.

\subsection{Combining GRBs and SNIa}

\begin{figure}\begin{center}
\resizebox{13cm}{12cm}{\includegraphics{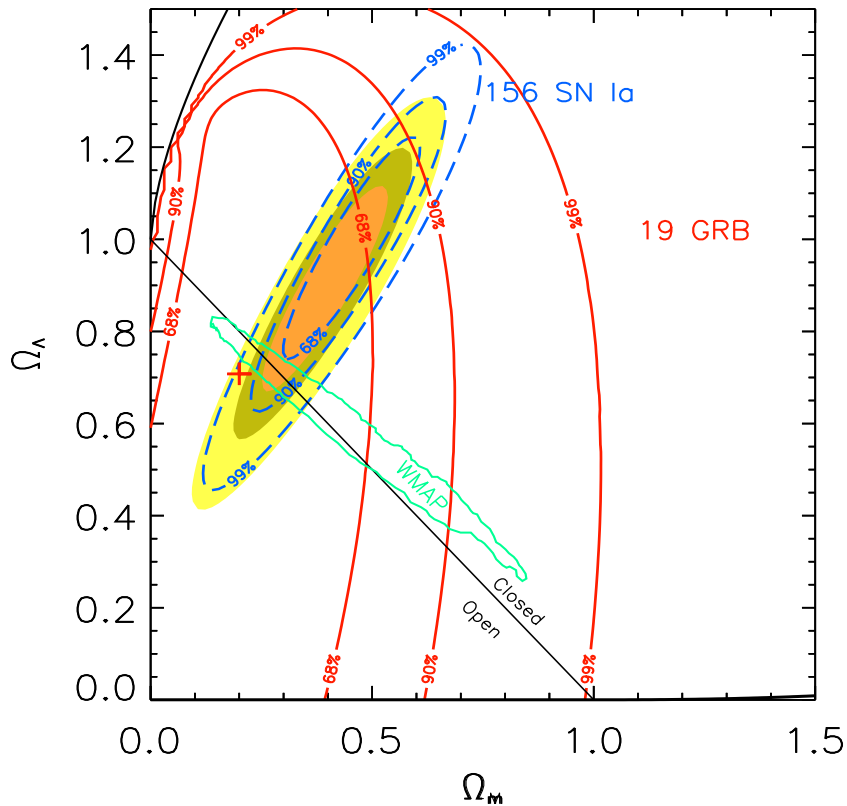}}
\caption{Constraints on the cosmological parameters $\Omega_{\rm M}$,
$\Omega_{\Lambda}$ obtained with the updated sample of 19 GRBs, in the
homogeneous density case (HM). The solid (red) contours, obtained with
the 19 GRBs alone, represent the 68.3\%, 90\% and 99\% confidence
regions. The center of these contours (red cross) corresponds to a
minimum $\chi^{2}=15.25/17$ dof and has $\Omega_{M}=0.23$ and
$\Omega_{\Lambda}=0.81$. The contours obtained with the 156 SN Ia of
the ``Gold'' sample of \cite{riess} are shown by the dashed
(blue) lines. The joint GRB+SN constraints are represented by the
shaded contours. We also show the 90\% confidence contours obtained
with the WMAP data (from \cite{spregel}).}\label{combo}\end{center}
\end{figure}

One important consequence of using GRBs as standard candles through
the $E_{\rm peak}$-$E_{\gamma}$ correlations or the empirical 3-D
$E_{\rm iso}$-$E_{\rm peak}$-$t_{\rm break}$ correlation is that they
can be combined with SNIa. With respect to SNIa, GRBs already extend
in a redshift range which is beyond the present (and future) SNIa
redshift limit (i.e. $z\sim1.7$ predicted for SNAP -
e.g. \cite{aldering}). Furthermore, GRBs are detected in the $\gamma$
ray band which is unaffected by dust extinction limitations. However,
SNIa are detected in large number also at very low redshifts and their
``stretching--luminosity'' correlation can be well
calibrated. Therefore, GRBs should be considered as complementary
to SNIa at very high redshisfts.

In Fig.~\ref{combo} we show the contours obtained with GRBs (in the HM
case) and we combine them with the constraints obtained for a large
sample of 156 SNIa (the ``gold'' sample of \cite{riess}). The
combined fit is clearly dominated by the large number of SNIa with
respect to GRBs (a factor 10 more numerous), however, the power of
GRBs is to make the joint contours more consistent with the
concordance model, also due to the different orientation of the
contours obtained by these two probes (see Sec.~1).

\subsection{Cosmological constraints with the $L_{\rm iso}$-$E_{\rm peak}$-$T_{0.45}$ correlation}

Finally, we present in Fig.~\ref{firmacosmo} the constraints
(\cite{firmani06a}) on the parameters $\Omega_{\rm M}$ and
$\Omega_{\Lambda}$ obtained with the ``prompt--emission'' correlation
$L_{\rm iso}$-$E_{\rm peak}$-$T_{0.45}$ (see Sec.~3.5). 

\begin{figure}\begin{center}
\resizebox{13cm}{12cm}{\includegraphics{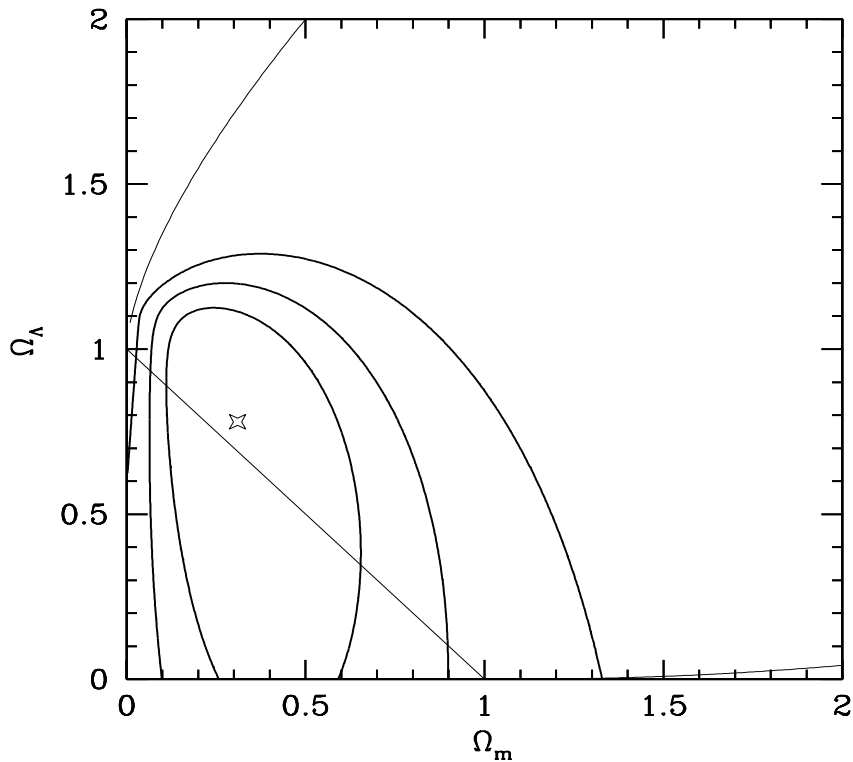}}
\caption{Contours at 68.3\%, 95.5\%, and 99.7\% CL's obtained from
projecting to the $\Omega_{\rm M}$, $\Omega_{\Lambda}$ plane the
$L_{\rm iso}$-$E_{\rm peak}$-$T_{0.45}$ relation that is obtained from
the fit of the GRB data at each value of the ($\Omega_{\rm
M}$,$\Omega_{\Lambda}$) pair. This plot shows that the $L_{\rm
iso}$-$E_{\rm peak}$-$T_{0.45}$ relation is sensitive to cosmology, so
that it may be used to discriminate cosmological parameters if an
optimal method to circumvent the circularity problem is used.  The
diagonal line corresponds to the flat geometry cosmology, the upper
curve is the loitering limit between Big Bang and no Big Bang models,
and the lower curve indicates the division between accelerating and
non-accelerating universes.}\label{aladinosc}\end{center}
\end{figure}

\begin{figure}\begin{center}
\resizebox{13cm}{12cm}{\includegraphics{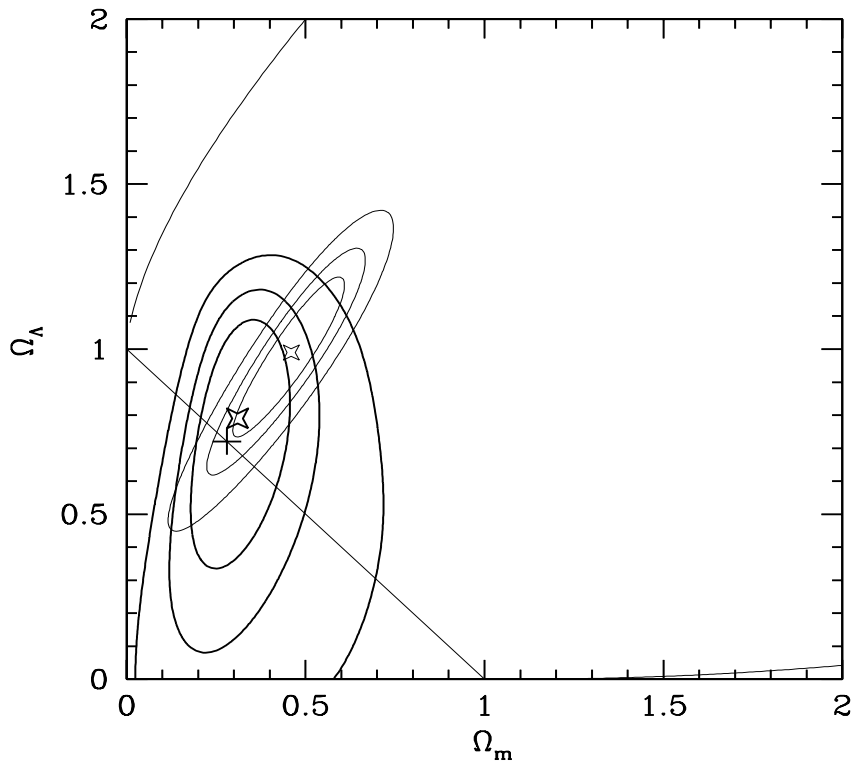}}
\caption{Constraints on the ($\Omega_{\rm M}$,$\Omega_{\Lambda}$)
plane for a $\Lambda$ cosmology from the GRB Hubble diagram using our
Bayesian method to circumvent the circularity problem and from the
gold set of SNIa of \cite{riess} (thin solid lines). The GRB
constraints are obtained with the $L_{\rm iso}$-$E_{\rm
peak}$-$T_{0.45}$ correlation.}\label{firmacosmo}\end{center}
\end{figure}

First we can apply the scatter method (I) to test if this correlation
is, indeed, sensitive to the cosmological parameters. We indeed show
in Fig.~\ref{aladinosc} smallest scatter is obtained for ($\Omega_{\rm
M}$,$\Omega_{\Lambda}$)=(0.3,0.7), i.e. very close to the concordance
model. 

By adopting the same Bayesian method described above, we note that the
constraints obtained with this new correlation with a sample of only
19 GRBs are tighter than those obtained with the $E_{\rm
peak}$-$E_{\gamma}$ correlation (Fig.~\ref{combo}).

The best cosmological model, indicated by the star symbol in
Fig.~\ref{firmacosmo}, corresponds to $\Omega_{\rm
M}=0.31^{+0.09}_{-0.08}$,$\Omega_{\Lambda}=0.80^{+0.20}_{-0.30}$
(1$\sigma$ uncertainty), i.e. vary close to the flat geometry
case. Indeed, if a flat geometry is assumed, we find $\Omega_{\rm
M}=0.29^{+0.08}_{-0.06}$ (\cite{firmani06a}).

\section{The future of GRB-cosmology}

The luminosity distance test requires large samples of sources
possibly distributed over a wide redshift range. This can be
understood from Fig.~\ref{sbisciole} which shows the topology of the
luminosity distance as a function of the cosmological parameters that
we want to constrain. In this case we consider $\Omega_{M}$ and
$\Omega_{\Lambda}$ as the two free parameters. The stripes in
Fig.~\ref{sbisciole} show the degeneracy of the $\Omega_{M}$ and
$\Omega_{\Lambda}$ parameters and how this degeneracy varies with the
redshift $z$. Each stripe represents, for a fixed redshift $z$, all
the possible combinations ($\Omega_{M}$,$\Omega_{\Lambda}$) which give
a luminosity distance equal within 10\%. By increasing the redshift
the stripes wound up due to the topology of the luminosity distance as
a function of these two parameters,
i.e. ($\Omega_{M}$,$\Omega_{\Lambda}$) (see \cite{ghisellini}). It is
clear from Fig.~\ref{sbisciole} that if we use a sample of sources
distributed over a wide range of redshifts we ``intersects''
differently oriented stripes and we end up with accurate constraints
on the two cosmological parameters. On the other hand a sample of
sources distributed in a very small redshift range would not break the
degeneracy of these two parameters. The requirement of a large number
of sources instead is a key ingredient to reduce the possible effect
of systematic errors.

The present open issues related to the use of GRBs in cosmologies are:
(1) the low number of events: the sample of GRBs which can be used to
constrain the cosmological parameters through the above correlations
are in fact 19 (to date); (2) the lack of a calibration sample, i.e.
the absence of GRBs at very low redshifts in the sample used for
cosmology, does not allow to calibrate the correlation and requires to
adopt a method to fit the cosmological parameters which avoids the
circularity problem; (3) the possible lensing effects which might
introduce a dispersion of the luminosities and energetics of GRBs at
very high redshifts. The latter point, however, if present, could be
easily recognized: if strong lensing by compact objects (if by
clusters they should be observed) is present, we should observe a
repetition of a similar structure (e.g. peak shape) within the same
light curve. Moreover, the lack of a theoretical interpretation
  of the physical nature of all these correlations represents a still
  open issue.

The increase of the number of bursts which can be used to measure the
cosmological parameters, and the possible calibration of the
correlations would greatly improve the constraints that are presently
obtained with few events and with a non--calibrated correlation.

In order to use GRBs as a cosmological tools, through the above
correlations, three fundamental parameters, i.e. $E_{\rm peak}$,
$E_{\gamma}$ and $t_{\rm jet}$, should be accurately measured.  This
requirement also applies to the case of the empirical correlation of
\cite{liang}. On the other hand the $L_{\rm iso}$-$E_{\rm
peak}$-$T_{0.45}$ does not require the knowledge of the afterglow
emission because it completely relies on the prompt emission
observables.

However, only a limited number of GRBs, i.e. 19 out of $\sim$ 70 with
measured $z$, can be used as standard candles.  Presently the most
critical observable is the peak energy $E_{\rm peak}$. In fact, the
limited energy range (15--150 keV) of BAT on board {\it Swift} allows
to constrain with only moderate accuracy the $E_{\rm peak}$ of
particularly bright--soft bursts. However, given the perspective of
the cosmological investigation through GRBs, it is worth exploring the
power of using GRBs as cosmological probes. The other still open issue
related to the use of GRBs as standard candles is the so called
``circularity problem'' (see Sec.~5).  

To the aim of calibrating this spectral--energy correlation, GRBs at
low redshift are required.  In fact for $z<0.1$ the difference in the
luminosity distance computed for different choices of the cosmological
models [for $\Omega_{\rm M},\Omega_{\Lambda}\in(0,1)$] is less than
8\%.  However, if (long) GRBs are produced by the death of massive
stars, they are expected to roughly follow the cosmic star formation
history (SFR) and we should expect that the rate of low redshift
events is quite small at $z<0.1$.  Indeed, a considerable number of
GRBs with $z>1$ should be collected in the next years by presently
flying instruments ({\it Swift} and {\it Hete--II}). At such large
redshifts, the cosmological models starts to play an important role.
However, if it will be possible to have a sufficient number of GRBs
with a similar redshift, it might still be possible to calibrate the
slope of these correlations also with high redshift GRBs (see
Sec.~7.3).

\subsection{The simulation}

In order to fully appreciate the potential use of GRBs for the
cosmological investigation, we simulate a sample of bursts comparable
in number to the ``Gold'' sample of 156 SN Ia. Similar simulations
have been presented before (e.g. \cite{xu},\cite{liang}). However,
different assumptions can be made on the properties of the simulated
sample and the results are clearly dependent on these assumptions. In
particular, simulations based on the observed parameters (\cite{xu})
strongly depend on the almost unknown selection effects which shape
the observed distributions.

\begin{figure}\begin{center}
\resizebox{13cm}{12cm}{\includegraphics{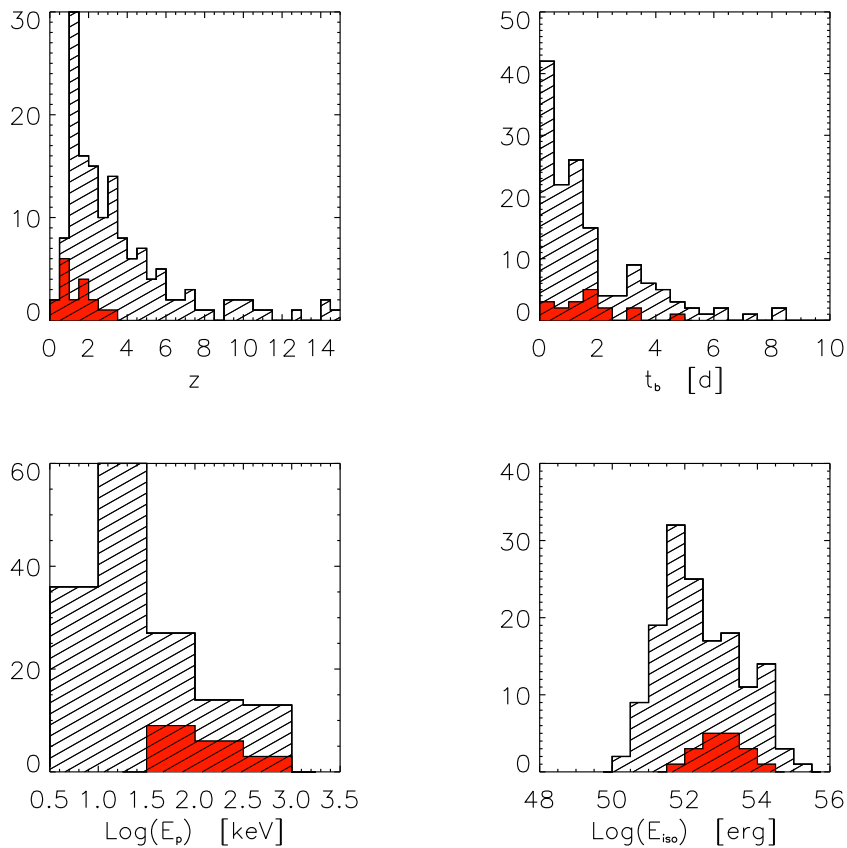}}
\caption{Distributions of redshift $z$, jet break time $t_{b}$,
observed peak energy $E_{\rm peak}^{\rm obs}$ and isotropic equivalent
energy $E_{\rm iso}$ for the 150 simulated bursts (hatched
histograms). Also show (solid filled histogram) are the distributions
of the sample of 19 GRBs used to derive the $E_{\rm peak}-E_{\rm iso}$
correlation.}\label{simula}\end{center}
\end{figure}

We adopt here a method which makes use of the intrinsic properties of
GRBs as described by the $E_{\rm peak}$-$E_{\gamma}$ and the $E_{\rm
peak}$-$E_{\rm iso}$ correlations. We use the most updated version of
these correlations as found with the sample of 19 GRBs described in
the previous sections.

The assumptions of our simulation are:
\begin{itemize}
\item we assume that GRBs have an ``isotropic energy'' function which
is described by a powerlaw $N(E_{\rm iso})\propto E_{\rm
iso}^{\delta}$ for $E_{\rm iso,min}<E_{\rm iso}< E_{\rm iso,max}$
(e.g. \cite{firmani04}). Further we assume that GRBs follow the cosmic
star formation rate (SFR - as modeled in \cite{porciani} );
\item we use the $E_{\rm peak}$-$E_{\rm iso}$ correlation to derive
the peak energy $E_{\rm peak}$ and we model the scatter of the
simulated GRBs around the $E_{\rm peak}$-$E_{\rm iso}$ correlation
with a gaussian distribution with $\sigma=0.2$ (which corresponds to
the present scatter of the 19 GRBs around their best fit correlation);
\item we use the $E_{\rm peak}$-$E_{\gamma}$ correlation as found with
the 19 GRBs in the WM case to calculate $E_{\gamma}$ and we model the
scatter of the simulated GRBs around the $E_{\rm peak}$-$E_{\gamma}$
correlation with a gaussian distribution with $\sigma=0.08$ (as found
in Sec.~4);
\item we derive the jet opening angle $\theta_{\rm j}$ and the
corresponding jet break time $t_{\rm break}$;
\item we assume that the simulated GRB spectra are described by a Band
model spectrum (\cite{band93}) with typical low and high energy
spectral photon indices $\alpha=-1.0$ and $\beta=-2.5$ and require
that the simulated GRB fluence in the 2-400 keV energy band is above a
typical instrumental detection threshold $S\sim 10^{-7}$
erg/cm$^2$. This corresponds roughly to the present threshold of {\it
Hete-II} in the same energy band.
\end{itemize}

Following the procedure described above we built a sample of 150 GRBs
with the relevant parameters: $z$, $E_{\rm iso}$, $E_{\rm peak}$,
$t_{\rm break}$. The errors associated to these parameters are assumed
to be cosmology invariant and they are set to 10\%, 20\% and 20\% for
$E_{\rm iso}$, $E_{\rm peak}$, $t_{\rm break}$ respectively. Moreover,
we model the GRB intrinsic isotropic energy function with a powerlaw
with $\delta=-1.3$ between two limiting energies ($10^{49}$-$10^{55}$
erg).  This particular choice of parameters is due to the requirement
that the distributions of the relevant quantities (shown in
Fig.~\ref{simula}) of the simulated sample are consistent with the
same distributions for the present sample of 19 GRBs (red histograms
in Fig.~\ref{simula}). The sample is simulated in the standard
cosmology ($\Omega_{M}=0.3$ and $\Omega_{\Lambda}=h=0.7$). We compare
the distributions of $z,t_b,E_peak$ and $E_{iso}$ for the 150
simulated bursts with the same distributions of the 19 GRBs in
Fig.~\ref{simula}. We  note that by choosing a steeper energy
function we obtain a much larger number of XRF and XRR with respect to
normal GRBs.

\subsection{Cosmological constraints with the simulated sample}

\begin{figure}\begin{center}
\resizebox{13cm}{13cm}{\includegraphics{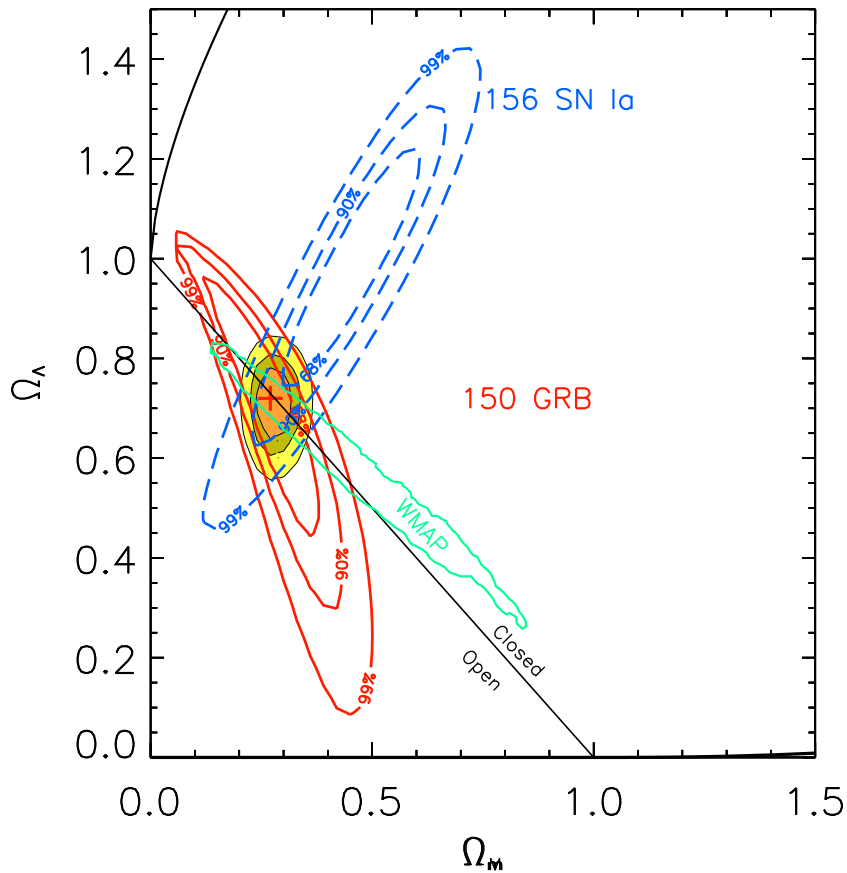}}
\caption{Wind density profile case. Constraints on the cosmological
parameters $\Omega_{\rm M}$,$\Omega_{\Lambda}$ obtained with the
sample of 150 GRBs simulated by assuming the $E_{\rm
peak}$-$E_{\gamma}$ correlations derived in the wind density
profile. The solid (red) contours, obtained with the 19 GRBs alone,
represent the 68.3\%, 90\% and 99\% confidence regions on the pair of
cosmological parameters. The contours obtained with the 156 SN Ia of
the ``Gold'' sample of \cite{riess} are shown by the dashed (blue)
lines. The joint GRB+SN constraints are represented by the shaded
contours.}\label{cosmo_150}
\end{center}
\end{figure}

The results obtained with the sample of 150 simulated GRBs are
presented in Fig. \ref{cosmo_150}: in this case the constraints are
comparable with those obtained with SNIa. By comparing the 1$\sigma$
contours of GRB alone from Fig.\ref{cosmo_150} to the same contours
(solid line) of Fig.~\ref{combo} (obtained with the 19 GRBs), we note
an improvement (of roughly a factor 10) with the sample of 150 bursts.

Fig.~\ref{cosmo_150} also shows the different orientation of the GRB
contours with respect to SN Ia due to the ``topology'' of the
luminosity distance as a function of the
$\Omega_{M}$-$\Omega_{\Lambda}$ parameters. Most GRBs of our simulated
sample are, in fact, at $z\sim 2$, and this explains the tilt of the
contours in the $\Omega$ plane. Clearly the contours obtained through
the simulated sample depends on the assumptions: in particular we have
no knowledge of the burst intrinsic energy function $N(E_{\rm
iso})$. However, if we accept the hypothesis to model it with a simple
powerlaw, we can change the slope and also include the possible effect
of the redshift evolution. We tested the dependence from these
assumptions of the constraints reported in Fig.~\ref{cosmo_150} and
found that, by assuming different $\delta$ values and a $(1+z)$
evolutionary factor, what changes is the redshift distribution of the
simulated sample and therefore the orientation of the GRB contours in
Fig.~\ref{cosmo_150}. The same happens if we adopt, for the same
choice of parameters reported above, a different SFR.

Further, we can also use the CMB priors. First we assume the cosmological 
constant model with the 2 CMB priors, i.e (i) $\Omega_{tot}=1$ and (ii)
$\Omega_{M}=0.14/h^{2}$. In this case the only free parameter is $h$
(or equivalently $\Omega_{M}$).We obtain the best fit values of
$\Omega_{M}=0.27\pm0.02$ and $\Omega_{\Lambda}=0.73\pm0.02$.

One of the major promises of the cosmological use of GRBs is related
to the possibility to study the nature of Dark Energy with such a
class of ``standard candles'' extending out to very large redshifts.
With the present sample of 19 GRBs we can explore the equation of
state (EOS) of DE, which can be parametrized in different ways. Given
the already considerably large dispersion of GRB redshifts
(i.e. between 0.168 to 3.2 for the 19 GRBs of our sample) we adopt the
parametrization proposed by {\cite{linder05} for the EOS of DE, i.e.
$P=w(z)\rho$, where:
\begin{equation}
w(z)= w_{0} + {w_{a}z \over 1+z}
\label{param}
\end{equation}
With  this assumption  the luminosity  distance, as  derived  from the
Friedman equations, is
\begin{eqnarray}\nonumber
d_{L}(z;\Omega_{M},w_{0},w_{a}) =  {c(1+z) \over H_{0}} \int_{0}^{z} dz 
[ \Omega_{M} (1+z)^{3}  &  \\
\,\,\,\,\,\,\,\,\,\,\,\,\,\,\,\,+ (1-\Omega_{\rm M})(1+z)^{3+3w_{0}+3w_{a}} 
\exp\left( -3w_{a} {z \over 1+z}\right)]^{-1/2} & 
\label{lumdist}
\end{eqnarray}
which depends on the ($w_{0}$,$w_{a}$) parameters. Note that
Eq. \ref{lumdist} is derived with the prior of a flat Universe.  

First, we can assume the CMB prior of a flat universe together with the
assumption of a non--evolving equation of state of the Dark Energy
(i.e. $w_{a}=0$). We show in Fig. 5 the contours obtained with the
sample of 150 simulated GRBs and compare with the same constraints
derived with the 156 SN Ia of the ``Gold'' sample. The constraints on
$w_{a}, w_{0}$ are reported in Fig. 6, assuming $\Omega_{\rm M}=0.3$.

\begin{figure}[htb]
\begin{minipage}[t]{8cm}
\resizebox{8cm}{8cm}{\includegraphics{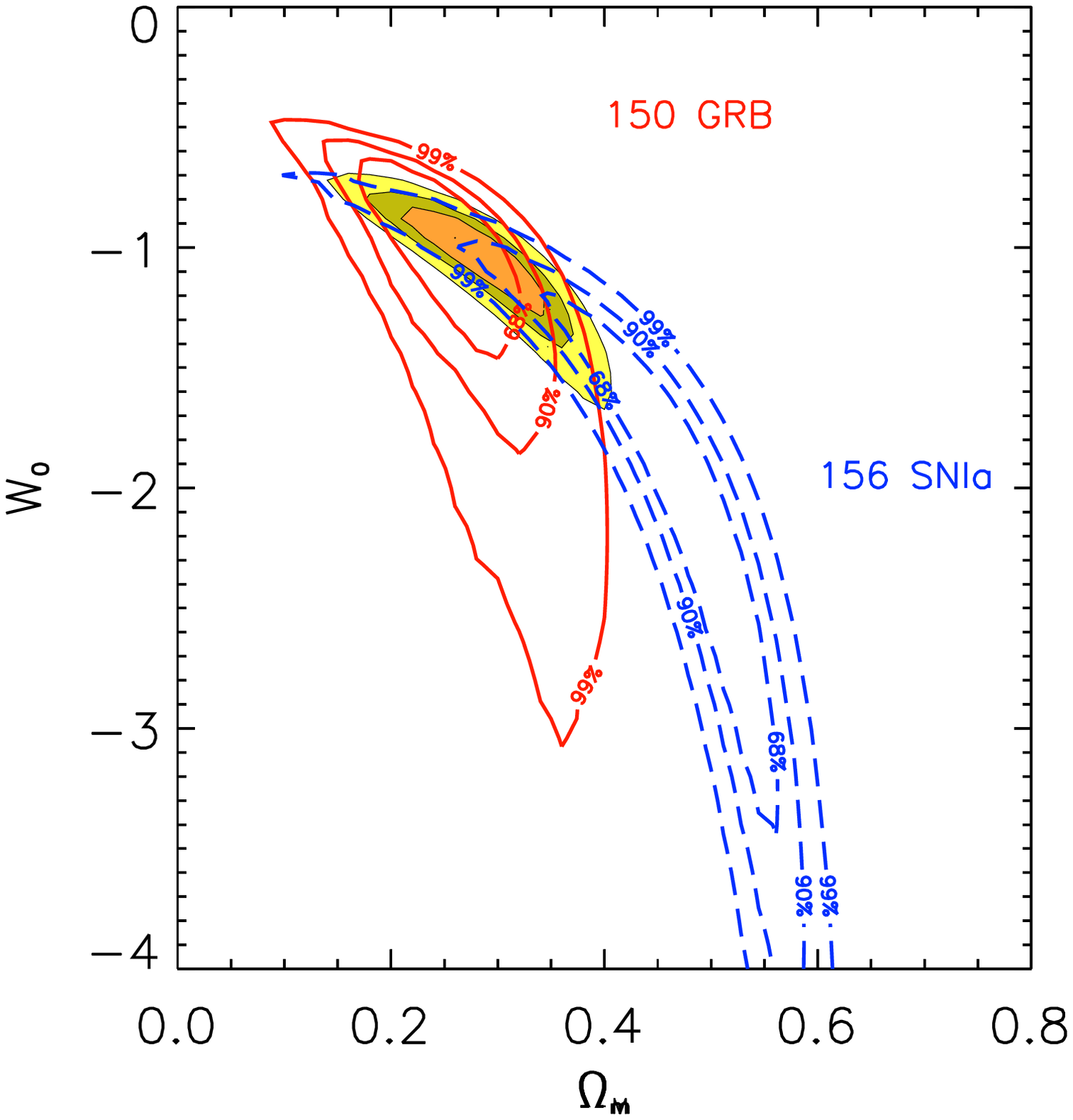}}
\end{minipage}\hfill
\begin{minipage}[t]{8cm}
\resizebox{8cm}{8cm}{\includegraphics{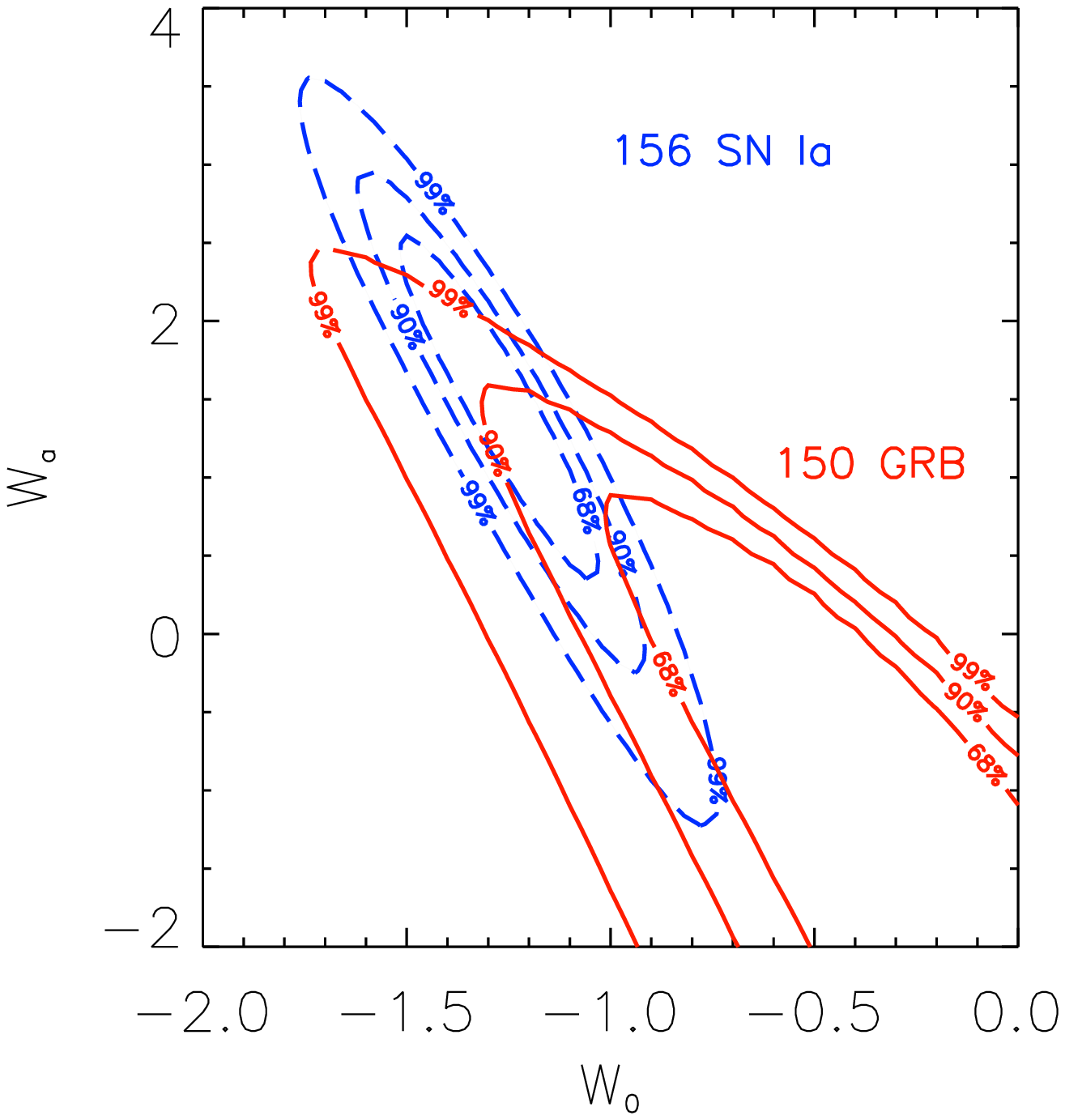}}
\end{minipage}
\caption[]{{\bf Left}: Constraints on the cosmological parameters
$w_{0}$, $w_{a}$ obtained with the 150 simulated GRBs (red
contours) compared with the same contours obtained with the 156 SNIa
of the ``Gold'' sample. A flat universe is assumed
($\Omega_{\rm tot}=1$). {\bf Right}: Constraints on the cosmological parameters
$w_{0}$, $\Omega_{\rm M}$ obtained with the 150 simulated GRBs (red
contours) compared with the same contours obtained with the 156 SNIa
of the ``Gold'' sample. A flat universe is assumed
($\Omega_{\rm tot}=1$). }
\label{edist2}
\end{figure}

\subsection{The calibration of the correlations}

The cosmological use of the $E_{\rm peak}=K\cdot E_{\gamma}^g$
correlation suffers from the so called ``circularity problem'' : this
means that both the slope $g$ and the normalization $K$ of the
correlation are cosmology dependent.

\begin{figure}\begin{center}
\resizebox{12cm}{14cm}{\includegraphics{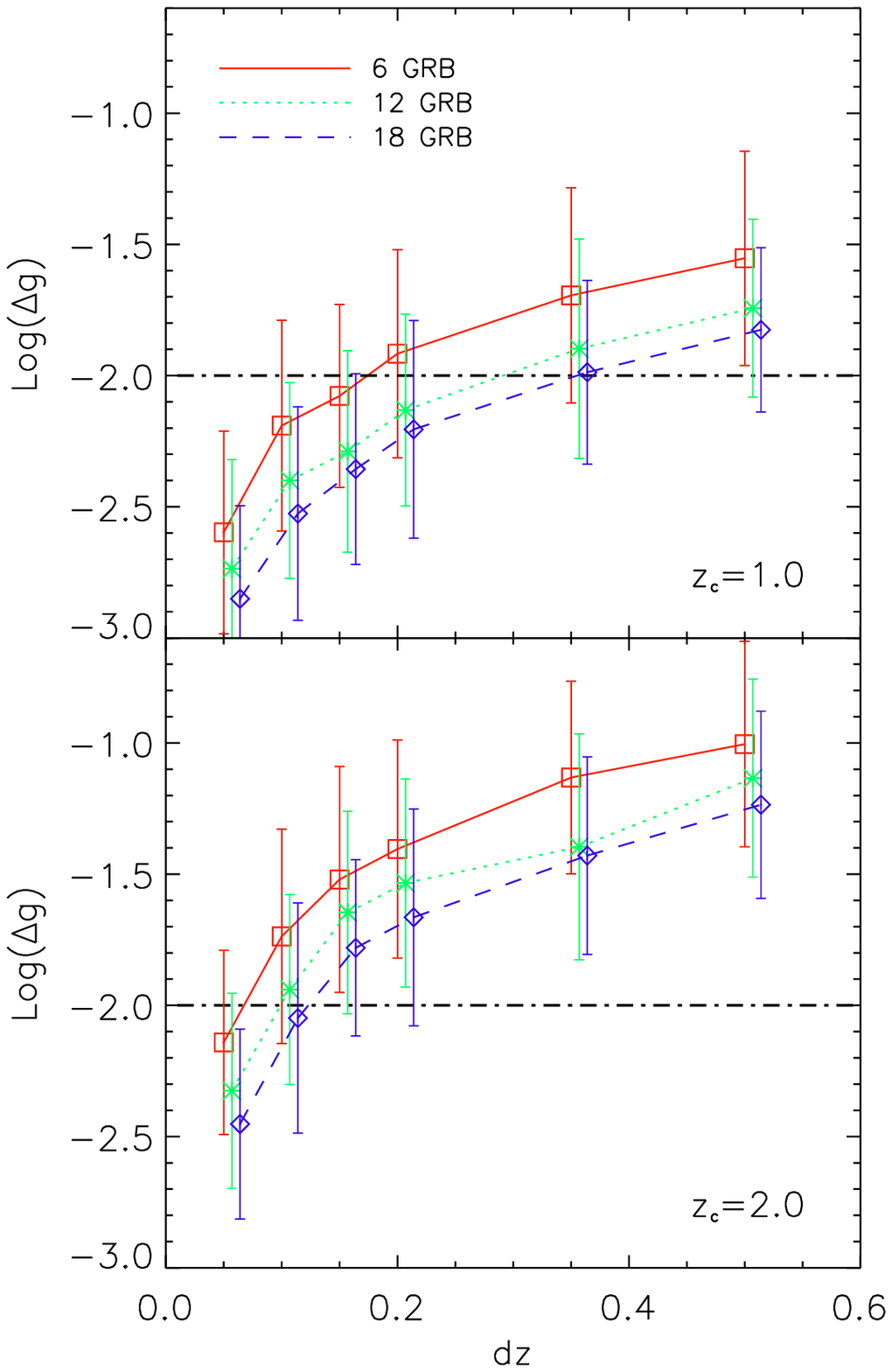}}
\caption{Calibration of the $E_{\rm peak}$-$E_{\gamma}^g$ correlation. 
For different samples of GRBs (6, 12, 18 - corresponding to
the solid, dotted and dashed lines respectively) we show the maximum
variation $\Delta g$ of the slope of the correlation for any cosmology
$\Omega\in(0,1.5)$ as a function of the redshift dispersion of the GRBs
$dz$. The dot dashed line represents the limit of variation of 1\% of
the slope of the correlation. Data points have been shifted along the 
abscissa for graphical purposes.}\label{calibra}\end{center}
\end{figure}

In principle this issue could be solved (a) by a large sample of
calibrators, i.e. low redshift GRBs for which the luminosity distance
$D_{L}$ is practically independent from the cosmological parameters,
or (b) by a convincing theoretical interpretation of the physical
nature of this correlation. In both cases the slope of the correlation
would be fixed.

Case (a) could be realized with 5--6 GRBs at $z<0.1$. However, if
(long) GRBs are produced by the core--collapse of massive stars, their
rate is mainly regulated by the cosmic SFR and, therefore, the
probability of detecting events at $z<0.1$ is small ($\sim
2\times10^{-5}$). This number should be convolved with the GRB
luminosity function: with the assumptions of our simulation we
estimate that $\sim 1.3$\% of the 150 GRBs should be at low redshifts
(i.e. $z<0.4$). Instead, we should expect to have more chances to
detect a considerable number (up to $\sim31$\%) of intermediate
redshift GRBs ($z\sim 1-2$) where the cosmic SFR peaks.

For this reason we explore the possibility to calibrate the
correlation using a sufficient number of GRBs within a small redshift
bin centered around {\it any} redshift. In fact, if we could have a
sample of GRBs all at the same redshift the slope of the $E_{\rm
peak}$-$E_{\gamma}$ correlation would be cosmology independent.  Our
objective is, therefore, to estimate {\it the minimum number of GRBs
($N$)} within a {\it redshift bin ($dz$)} centered around a certain
{\it redshift ($z_c$)} which are required to calibrate the
correlation.

In practice the method consists in fitting the correlation for every choice 
of $\Omega$ using a set of $N$ GRBs distributed in the interval $dz$ (centered 
around $z_c$). We consider the correlation to be calibrated (i.e. its slope to 
be cosmology independent) if the change of the slope $g$ is less than 1\%.

The free parameters of this test are the number of GRBs $N$, the
``redshift slice'' $dz$ and the central value of the redshift
distribution $z_c$.  By Monte Carlo technique we use the sample
simulated in Sec.~7.1 under the WM assumption to minimize the
variation of $\Delta g(\Omega;N,dz,z_c)$ over the
$\Omega_{M},\Omega_{\Lambda}\in(0,1.5)$ plane as a function of the
free parameters $(N,dz,z_c)$.

We tested different values of $z_{c}$ and different redshift
dispersions $dz\in(0.05,0.5)$. We required a minimum number of 6 GRBs
to fit the correlation in order to have at least 4 degrees of freedom.
We report our results in Fig.~\ref{calibra}. We show the variation of
$\Delta g$ as a function of $dz$ for different samples (6, 12, 18 GRBs
- solid, dotted and dashed curves in Fig.~\ref{calibra}). The error
bars show the width of the distribution of the simulation results and
not the uncertainty on the average value. At any redshift the fewer
the number $N$ of GRBs the larger the change of $\Delta g$ (for the
same $dz$) because the correlation is less constrained. The dependence
from $z_c$ is instead different: for larger $z_c$ we require a smaller
bin $dz$ to keep $\Delta g$ small.

From the curves reported in Fig.~\ref{calibra} we can conclude that
already 12 GRBs with $z\in(0.9,1.1)$ might be used to calibrate the
slope of the $E_{\rm peak}$-$E_{\gamma}$ correlation.  At redshift
$z_c=2$ instead we require a smaller redshift bin
i.e. $z\in(1.95,1.05)$.  We find that $N=12$ GRBs with
$z\in(0.45,0.75)$ can be used to achieve the same 1\% precision in the
calibration. However, one key ingredient is that the GRBs used to
calibrate the correlation do not have the same peak energy otherwise
they would collapse in one point in the $E_{\rm peak}$-$E_{\gamma}$
plane.  Within the present sample of 19 GRBs there are only 4 GRBs
within the redshift interval 0.4--0.8 (i.e. 050525, 041006, 020405 and
051022) and 2 of these (050525 and 041006) have a very similar $E_{\rm
peak}$.

We conclude stressing that the three correlations that can be used to
constrain the cosmological parameters, i.e. the $E_{\rm
peak}$-$E_{\gamma}$ (either in the HM and WM case -
\cite{ghirlanda04,ghirlanda04a,firmani,nava,ghirlanda06}), the
empirical $E_{\rm iso}$--$E_{\rm peak}$--$t_{\rm break}$ correlation
(\cite{liang}) and the $L_{\rm iso}$-$E_{\rm peak}$-$T_{0.45}$
correlation (\cite{firmani06,firmani06a}), all involve the peak energy
$E_{\rm peak}$ of the GRB prompt emission spectrum. This observable
can be properly measured with a detector operating over a wide energy
range extending up to few MeV as it could be conceived by future
missions dedicated to collect and study the prompt and afterglow
properties of GRB to be used as standard candles for cosmology
(\cite{lamb05}).

\section{Acknowledgements}

We would like to thank F. Tavecchio, D. Lazzati, L. Nava and
M. Nardini for stimulating discussions. We are deeply grateful to
Annalisa Celotti for years of fruitful and enjoying collaboration. The
bibliography is clearly incomplete and we apologize for any minor or
major missing reference.

\end{document}